\documentclass[useAMS,usenatbib]{mn2e}

\def \den{g.cm$^{-3}$}
\usepackage{graphicx}

\title[Axisymmetric smoothed particle hydrodynamics with self-gravity]{Axisymmetric smoothed particle hydrodynamics with self-gravity}
\author[D. Garc\'\i a-Senz et al.]{D. Garc\'\i a-Senz$^{1,2}$\thanks{E-mail: domingo.garcia@upc.edu}, 
A. Rela\~{n}o$^{1}$\thanks{E-mail: antonio.relano@upc.edu}, 
 R.M. Cabez\'on$^{1}$\thanks{E-mail:ruben.cabezon@upc.edu} and E. Bravo$^{1,2} $\thanks{E-mail: eduardo.bravo@upc.edu}\\ 
$^{1}$Departament de F\'\i sica i Enginyeria Nuclear (UPC), Jordi Girona 3, 
Barcelona, 08034, Spain\\
$^{2}$Institut d'Estudis Espacials de Catalunya, Gran Capit\`a 2-4, Barcelona, 08034, Spain} 
\begin{document}
\date{Accepted for publication in MNRAS}

\pagerange{\pageref{firstpage}--\pageref{lastpage}} \pubyear{2008}

\maketitle

\label{firstpage}

\begin{abstract}

The axisymmetric form of the hydrodynamic equations within the
smoothed particle hydrodynamics (SPH) formalism is presented  and
checked
using idealized scenarios taken from astrophysics (free fall collapse,  
implosion and further 
pulsation of a sun-like star), gas dynamics (wall heating problem, collision 
of two streams of gas) and inertial
confinement fusion (ICF, -ablative implosion of a small capsule-). New material concerning the standard SPH
formalism is given. That includes the numerical handling of those mass
points which move close to the singularity axis,  more accurate
expressions for the artificial viscosity and 
the heat conduction term and an easy way to incorporate self-gravity in the
simulations. The algorithm developed to compute gravity does not
rely in any sort of grid, leading to a numerical scheme totally
compatible with the lagrangian nature of the SPH equations. 

\end{abstract}


\section{Introduction}

Since the smoothed-particle hydrodynamics method was proposed by  
\citet{gm77}~and \citet{lu77}
thirty years ago it has become a powerful and  very successful method, which  
is routinely used worldwide 
to model systems with complicated geometries.  
In particular the combination of the SPH method 
 with hierarchical algorithms, addressed to efficiently organize 
 data and calculate  gravity, have led to a robust and reliable schemes   
able to handle complex geometries in three dimensions.
In spite of these achievements little effort has been put in the 
development of axisymmetric applications of the method.  
However,  
a large body of interesting applications has to do with 
systems which have axisymmetrical geometry such as accretion discs, 
rotating stars or explosive phenomena (i.e. novae or supernovae events) 
provided 
the ignition takes place in a point-like region at the symmetry axis. The interaction between the gas ejected during a supernova explosion and the 
circumstellar matter has been extensively simulated using cylindric 
coordinates 
either with eulerian \citep{bl96} or SPH \citep{ve06} codes.  In 
Inertial Confinement Fusion (ICF) studies the collapse of the 
deuterium capsule induced by laser ablation can be also approximated 
using two-dimensional hydrodynamics.
Axisymmetrical hydrocodes are also useful to conduct resolution studies of three-dimensional codes just 
by imposing an initial configuration with the adequate symmetry. In these cases a 
comparison between the results obtained using the 3D-hydrocode and the 
equivalent, 
although usually better resolved, 2D simulation could  help in numerical 
convergence studies. 
Several formulations of different complexity have been proposed to handle 
with the axisymmetric version of SPH. While the first algorithms were 
in general very crude, i.e. \citet{he92} and references therein, 
neglecting  
the so called hoop-stress terms, there were also 
more consistent SPH approximations,   
\citep{pe93}. Recently more complete formulations
 were given by \citet{br03} and by \citet{om06}. In 
particular the scheme devised by Omang et al. includes not only the 
hoop-stress terms but also a consistent treatment of the singularity line 
defined by the 
symmetry axis. The algorithm proposed by these authors relies in the use of interpolative 
kernels especially adapted to the cylindrical geometry. Although the resulting 
scheme is robust, being able to successfully pass several test cases, it 
has the numerical inconvenient that an elliptical integral has to be solved 
numerically at each integration step for each particle. 

In this paper we propose a new formulation of the axisymmetric SPH technique 
which preserves many of the virtues stated by \citet{br03} and  \citet{om06},  
being also  able to handle the singularity axis in an efficient way.
In our scheme there is no necessity  
to modify the interpolating kernel (we use the cubic spline polynomial). Instead, taking advantage of the symmetry properties across the Z-axis, we found 
several analytical correction factors to the physical magnitudes. 
These correction factors only affect  
particles moving at a distance from the Z-axis lesser than $2h$, being $h$~the smoothing-length 
parameter. For the remaining particles the SPH equations are identical to 
those given in \citet{br03}. In addition, we also give expressions 
for the artificial viscosity and a new heat conduction algorithm,  
which takes into account the hoop-stress contribution. 
Finally, we  propose an original method to handle gravity within the SPH 
paradigm which is consistent with the gridless nature of that particle method. 
The text is organized as follows: the basic fluid equations written in 
axisymmetric coordinates, and corrected from axis effects when necessary,  
are provided in Section 2. In Section 3 we add useful physics to these 
equations 
consisting of an artificial viscosity term to handle shock waves (Section 3.1), 
an expression for thermal conduction (Section 3.2) and an algorithm  
addressed to calculate self-gravity  within the axisymmetric hypothesis 
(Section 3.3). 
Section 4 is devoted to describe and discuss five test cases aimed at 
validating the 
proposed scheme. Finally the main conclusions of our work as well as
 some comments about the 
limitations of the developed scheme  
 and prospects for the future are presented in 
Section 5.

\section[]{Fluid equations in the axisymmetric approach.}

An elegant approach to the axisymmetric Euler fluid equations was given 
by \citet{br03} who derived the SPH form of these basic equations using the 
minimal action principle (see Monaghan 2005 and references therein for the history of variational SPH). The resulting expressions for mass, momentum and 
energy conservation written in the cylindrical coordinate 
system ${\mathbf s}= (r,z)$
 are:

\begin{eqnarray}
\eta_i=\sum_{j=1}^N m_j W_{ij}
\end{eqnarray}

\begin{eqnarray}
\cases{\ddot{r}=2\pi\frac{P_i}{\eta_i}-2\pi\sum\limits_{j=1}^N m_j \left(\frac{P_i r_i}{\eta_i^2}+\frac{P_j r_j}{\eta_j^2}\right)\frac{\partial W_{ij}}{\partial r_i}\cr\cr
\ddot{z}=-2\pi\sum\limits_{j=1}^N m_j \left(\frac{P_i r_i}{\eta_i^2}+\frac{P_j r_j}{\eta_j^2}\right)\frac{\partial W_{ij}}{\partial z_i}\cr\cr}
\end{eqnarray}

\begin{eqnarray}
\frac{du_i}{dt}=-2\pi\frac{P_i~{v_{r_i}}}{\eta_i}+
 2\pi\frac{P_i~r_i}{\eta_i^2}\sum_{j=1}^N m_j (\mathbf{v_i-v_j})\cdot \mathbf{D}_i~W_{ij}
\end{eqnarray}

\noindent
where $\eta_i=2\pi r_i\rho_i$~is the  
two-dimensional density of the i-particle, $\mathbf {v}=(v_r,v_z)$~its 
velocity,   
$W_{ij}=W_{ij}(\frac{\vert\mathbf{s_i-s_j\vert}}{h_i})$~the  
interpolating kernel and   
the remaining symbols have their usual meaning. The differential operator 
${\mathbf D}= (\frac{\partial}{\partial r},\frac{\partial}{\partial z})$ 
  is a cartesian operator written in the $(r,z)$~ plane of cylindric 
coordinates.  
The first terms on the right 
in the r-component of equation (2) and equation (3) are called the hoop-stress terms. They are geometrical 
terms which arise from the nabla operator written in cylindrical coordinates.
The smoothing-length parameter, $h$, is 
usually taken as the local resolution of the SPH. Equations (1), (2) and (3) 
along with the adequate equation of state (EOS) 
are the starting point to carry out numerical simulations of fluid systems 
with axisymmetric geometry. As boundary conditions we consider reflective 
ghost particles across the Z-axis and open flow at the outer limits of the 
system.  
For the $i$-particle~with coordinates $(r_i, z_i)$~and velocity 
$(\dot r_i, \dot z_i)$~it is defined its reflective $k$-particle by taking 
$(r_k,z_k)=(-r_i, z_i)$, $(\dot r_k, \dot z_k)=(-\dot r_i, \dot z_i)$~and 
$m_k=m_i$. 
Thus position, velocity as well as other magnitudes such as density or internal 
energy of reflective particles are updated at each step not using the SPH 
equations but from the evolution of real particles.  
Even though the use of reflective boundary conditions is not strictly necessary in axisymmetric geometry it is useful to correctly represent the density
and its derivatives near to the singularity axis.  

\subsection{Correction terms close to the singularity axis}

One of the major causes of inaccuracy of axisymmetric SPH is the treatment  
of particles that get close to the symmetry axis. Unlike  
spherically symmetric systems where there is only one singular point, just at 
the center,   
here there is a singular line at $r=0$. A good treatment of  
particles moving close to the Z-axis is especially relevant for implosions 
such as the collapse of a self-gravitating system or in inertial confinement 
fusion studies. As far as we know only \citet{om06} have consistently addressed this problem. In that paper the interpolation kernel is modified 
according to the particular geometry of the system, spherical or cylindrical. 
The resulting scheme does not have singularity problems when the particles 
approach the axis. However the resulting kernel does not have an analytical 
expression and must be calculated at any step for each particle using a 
numerical integration. Recently,  useful fitting formulae were proposed by 
\cite{om07}~but still involving a large number of operations which slows the 
calculation.

An alternative way to handle with the symmetry axis, without modifying the 
basic 
SPH scheme given above, is to calculate correction terms 
to equations (1), (2) and (3), which become significant only close to the Z-axis. These correction 
factors arise because of the limited capability of standard kernels to 
interpolate accurately 
non linear functions. In the particular case of the two-dimensional density, 
$\eta=2\pi\vert r\vert\rho$, its profile is not longer linear in the 
axis neighbourhoods due to the 
presence of reflective particles. Usually the errors are small and the interpolation is precise to 
second order in $h$. Unfortunately, close to the symmetry axis errors grow 
and 
density and other physical magnitudes are not well reproduced, as it  
will be shown below.  
The 
detailed derivation of these corrections are given in Appendix A, leading to 
the following equations:

\begin{eqnarray}
\widehat{\eta}_i=\sum_{j=1}^N m_j W_{ij}\times f_1^i=
   \eta_i\times f_1^i
\end{eqnarray}

\noindent
where $\widehat{\eta}_i$~is the new, improved, two-dimensional density 
and $f_1^i$~is
a correction factor which, for the cubic-spline kernel, reads (Appendix A):

\begin{eqnarray}
f_1^i=
\cases
{\left[\frac{7}{15}\zeta_i^{-1}+\frac{2}{3}
\zeta_i-
\frac{1}{6}\zeta_i^3+\frac{1}{20}\zeta_i^4\right]^{-1}\qquad 0<\zeta_i < 1  
\cr\cr 
 \left[\frac{8}{15}\zeta_i^{-1}-\frac{1}{3}+\frac{4}{3}
\zeta_i-
\frac{2}{3}\zeta_i^2+\frac{1}{6}\zeta_i^3-\frac{1}{60}\zeta_i^4\right]^{-1}\cr 
\qquad\qquad\qquad\qquad {\mathrm for} \qquad\qquad 1<\zeta_i<2       \cr\cr 
 1 \qquad\qquad\qquad\qquad \zeta_i >2 \cr\cr}   
\end{eqnarray}

\noindent
being $\zeta_i=r_i/h_i$. A plot of $f_1(\zeta)$~and its first derivative is given in 
Fig.~1. Hereafter a hat will be placed over any corrected, and 
therefore, "true" magnitude. 

 \begin{figure}
 \center
 \includegraphics[scale=0.40]{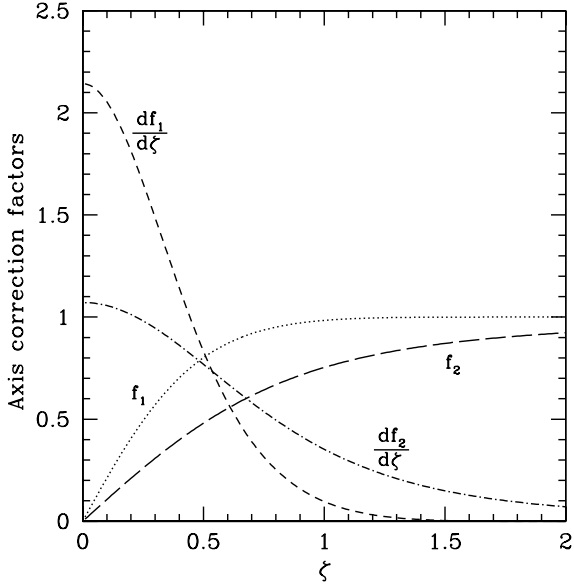}
 \caption{Plot of the correction factors $f_1$~and $f_2$~and their derivatives 
 as a function of $\zeta=r/h$. The cubic spline interpolator kernel was assumed to 
compute $f_1$~and $f_2$~(Appendix A).
  }\label{fig1}
\end{figure}

At this point it is useful to introduce a couple of algebraic rules: 

\begin{equation}
\widehat A_i=\sum_j m_j\frac{A_j}{\widehat\eta_j}W_{ij}
\end{equation}

\begin{equation}
\widehat{\eta A}_i=f_1^i\sum_j m_j A_j W_{ij}
\end{equation}

\noindent which are valid as long as, close to the axis, the magnitude $A$~has a weak dependence 
in the 
r-coordinate. In particular, setting $A=1$~
in equation (7) leads to equation (4).

To find the influence of the above correction factor in the 
momentum equation, equation (2), its is better to write the  
components of that  
equation in differential form. For the radial component we have:

\begin{eqnarray}
\ddot{r}=-\frac{1}{\rho}\frac{\partial P}{\partial r}=-\frac{2\pi r}{\widehat\eta}\frac{\partial P}{\partial r}=2\pi\frac{P}{\widehat\eta}-\frac{2\pi rP}{\widehat\eta^2}\frac{\partial\widehat\eta}{\partial r}-
    \frac{\partial}{\partial r}\left(\frac{2\pi rP}{\widehat\eta}\right)
\end{eqnarray}

\noindent
When $r\rightarrow 0$~the particle approaches the Z-axis 
where symmetry enforces  
the last term on the right to vanish for spherically symmetric kernels. The 
first term on the right is 
the hoop-stress term which should be exactly balanced by the 
central term when $r\rightarrow 0$, because the acceleration should  be zero 
at the 
symmetry axis. Nevertheless, this does not happen
unless the 
correction factor $f_1(\zeta)$~is taken into account during the calculation of the 
gradient of 
$\widehat\eta$: 

\begin{eqnarray}
\frac{\partial\widehat\eta}{\partial r}=\frac{\partial (f_1(\zeta)\eta)}{\partial r}=f_1(\zeta)\frac{d\eta}{dr}+
\eta\frac{\partial f_1(\zeta)}{\partial r}
\end{eqnarray}

Similarly, for the Z-component in the momentum equation:

\begin{eqnarray}
\ddot{z}=-\frac{1}{\rho}\frac{\partial P}{\partial z}=-\frac{2\pi r}{\widehat\eta}\frac{\partial P}{\partial z}=-\frac{2\pi rP}{\widehat\eta^2}\frac{\partial \widehat\eta}{\partial z}-
    \frac{\partial}{\partial z}\left(\frac{2\pi rP}{\widehat\eta}\right)
\end{eqnarray}

\noindent
where now we simply have:

\begin{eqnarray}
 \frac{\partial\widehat\eta}{\partial z}=f_1(\zeta)\frac{\partial \eta}{\partial z}
\end{eqnarray}

The resulting momentum components in discrete SPH form can be obtained from 
equations (6) to (11):

\begin{eqnarray}
\nonumber
 \ddot{r}_i=2\pi\frac{P_i}{\widehat\eta_i}-2\pi \sum_{j=1}^N\left[ m_j\left(
 \frac{P_i r_i}{\widehat\eta_i^2}\times f_1^i(\zeta_i)+\frac{P_j r_j}{\widehat\eta_j^2}
\right)\frac{\partial W_{ij}}{\partial r_i}\right]-\\ 
\left(\frac{2\pi P_i~r_i}{\widehat\eta_i\times f_1^i}\right)
~\frac{df_1^i(\zeta_i)}{dr_i}\qquad\qquad
\end{eqnarray}

\begin{eqnarray}
\ddot{z}_i=-2\pi \sum_{j=1}^N m_j\left(
 \frac{P_i r_i}{\widehat{\eta_i}^2}\times f_1^i(\zeta_i)+\frac{P_j r_j}{\widehat{\eta_j}^2}
\right)\frac{\partial W_{ij}}{\partial z_i} 
\end{eqnarray}

\noindent
The value of pressure in equations (12) and (13) must be computed through the 
EOS using the 
'corrected' 3D-density, $\widehat\rho=\widehat\eta/(2\pi r)$. 

Note that close to the axis the summatories in equations (12) and (13)  are
 non symmetric with  
respect any pair of 
particles. Still the r-component of the momentum is conserved, even for those 
particles with $r<2h$, because of the imposed reflective boundary conditions. 
In the 
Z-axis the momentum is not exactly conserved within the band defined by 
$r<2h$. In general such loss is small, affecting a tiny amount of mass. In  
most of the tests described below 
total momentum was very well preserved.  
 
A similar approach can be worked out to improve the energy equation:  

\begin{eqnarray}
\frac{du}{dt}=-2\pi\frac{P}{\widehat\eta}~v_{r}+2\pi
  \frac{P r}{\widehat{\eta}^2}\frac{d\widehat{\eta}}{dt}
\end{eqnarray}

\noindent where:

\begin{eqnarray}
\frac{d\widehat\eta}{dt}=\left[v_r\frac{\partial\widehat\eta}{\partial r}-
 \frac{\partial\widehat{(\eta v_r)}}{\partial r}\right]+\left[v_{z}\frac{\partial\widehat{\eta}}{\partial z}-\frac{\partial\widehat{(\eta v_z)}}{\partial z}\right]
\end{eqnarray}
  
\noindent
being $\widehat\eta=\eta\times f_1$;~$\widehat{(\eta v_z)}=(\eta v_z)\times f_1$~and 
$\widehat{(\eta v_r)}=(\eta v_r)\times f_2$. The factor  $f_2$~is (Appendix A):

\begin{eqnarray}
f_2^i=
\cases
{\left[\frac{14}{15}\zeta_i^{-1}+\frac{4}{9}
\zeta_i-
\frac{1}{15}\zeta_i^3+\frac{1}{60}\zeta_i^4\right]^{-1}\qquad 0<\zeta_i < 1  
\cr\cr 
 \big[-\frac{1}{45}\zeta_i^{-2}+\frac{16}{15}\zeta_i^{-1}-
\frac{1}{3}+\frac{8}{9}\zeta_i-\frac{1}{3}\zeta_i^2+\frac{1}{15}\zeta_i^3-\cr\cr
\qquad \frac{1}{180}\zeta_i^4 \big]^{-1}\qquad\qquad\qquad  
  \mathrm {for}\qquad  1<\zeta_i<2       \cr\cr 
 1 \qquad\qquad\qquad\qquad \zeta_i >2 \cr\cr}   
\end{eqnarray}

The resulting energy equation in discrete SPH form is:

\begin{eqnarray}
\frac{du_i}{dt}=-2\pi\frac{P_i}{\widehat\eta_i}~v_{r_i}+2\pi
  \frac{P_i r_i}{\widehat{\eta_i}^2}\frac{d\widehat{\eta_i}}{dt}
\end{eqnarray}

\noindent
where:

\begin{eqnarray}
\frac{d\widehat{\eta_i}}{dt}=
\sum_{j=1}^N m_j\left(f_1^i v_{r_i}-f_2^i v_{r_j}\right)
\frac{\partial W_{ij}}{\partial r_i}+\cr\cr
\sum_{j=1}^N m_j \left(\frac{\partial f_1^i}{\partial r_i}v_{r_i} 
-\frac{\partial f_2^i}{\partial r_i} v_{r_j}\right) W_{ij}\cr\cr  
+f_1^i\sum_{j=1}^N m_j\left(
v_{z_i}-v_{z_j}\right)\frac{\partial W_{ij}}{\partial z_i}
\end{eqnarray}

Note that $f_1^i$, $f_2^i$~and their derivatives are only function of the current radial coordinate 
of the particle and its smoothing length. Thus, they can easily be computed and stored in 
 a vector without introducing any significant computational overload. 
 Examples about the use of the axisymmetric fluid equations with axis corrections, equations (4), (12), (13) and (17) in practical situations  
will be provided and discussed in Section 4.

\section[]{Adding physics: Shocks, Thermal conduction and gravity}

\subsection{Artificial viscosity}

As in three dimensions the treatment of shocks in the axisymmetric 
approach also relies in the artificial viscosity formalism. Nonetheless, there 
are a variety of artificial viscosity algorithms suited for SPH to choose at. 
We have adapted the standard recipe by 
\citet{mo83} to the peculiarities of the 
2D-axisymmetric hydrodynamics.  
In that approach the 
artificial viscosity gives rise to a viscous bulk and shear pressure which 
is added to the 
normal gas pressure only in those regions where the particles collide (in 
the SPH sense\footnote{In the SPH method the so called particles could be looked as  finite spheres of radii $2h$~}). In three dimensions it is defined a  
magnitude, $\Pi_{ij}$:

\begin{equation}
\Pi_{ij}^{3D}=\cases
{\frac{-\alpha_0\bar c_{ij}\mu_{ij}+\beta_0{\mu^2_{ij}}}
{\bar\rho_{ij}}\qquad\qquad\qquad \mathbf {v_{ij}}\cdot \mathbf {r_{ij}}<0\cr\cr
0\qquad\qquad\qquad\qquad\qquad\quad \mathbf {v_{ij}}\cdot \mathbf {r_{ij}}>0\cr}
\label{eq:eq24}
\end{equation}

\noindent
which is closely related to the viscous pressure. In equation (19) $\alpha_0$~and $\beta_0$~are constants of the order of unity, 
$\bar \rho_{ij}$~and $\bar c_{ij}$~
are the average of density and sound speed of $i$~and $j$- particles, and 
$\mu_{ij}$~is:

\begin{equation}
\mu_{ij}=\frac{h_{ij}~\mathbf{v}_{ij}\cdot~\mathbf{r}_{ij}}{r^2_{ij}+\nu^2}
\label{eq:eq25}
\end{equation}
 
\noindent
here $\mathbf v_{ij}=(\mathbf v_i-\mathbf v_j)$, $\mathbf{r}_{ij}=(\mathbf r_i-\mathbf r_j)$~and  $\nu=0.1h$~avoids divergences when $r_{ij}\rightarrow 0$. 
The scalar magnitude $\Pi_{ij}$~has a linear and a quadratic term. The linear 
component mimics bulk and shear viscosity of fluids whereas the quadratic one 
is important to avoid particle interpenetration in strong shocks. 

The easiest way to write the contribution of the artificial viscosity to the 
momentum and energy equations in an axisymmetric code is by changing 
the mass of the particles according to their distance to the Z-axis: 
$m\longrightarrow  \frac{m}{2\pi r}$. The viscous acceleration becomes:

\begin{equation}
\mathbf a_i^{vis}=\sum_{j=1}^N \frac{m_j}{2\pi\bar r_{ij}}   
\Pi_{ij}^{3D}\mathbf {D}_i W_{ij}
\end{equation}

Taking  $\bar\rho_{ij}=
\bar\eta_{ij}/(2\pi \bar r_{ij})$~in equation (19) the explicit 
dependence on $r_{ij}$~
in the viscous acceleration formula is removed,

\begin{equation}
\mathbf a_i^{vis}=\sum_{j=1}^N m_j   
\Pi_{ij}^{(1)}\mathbf {D}_i W_{ij}
\end{equation}

\noindent
where $\Pi_{ij}^{(1)}$~is:

\begin{equation}
\Pi_{ij}^{(1)}=\cases
{\frac{-\alpha_0\bar c_{ij}\mu_{ij}+
\beta_0~\mu^2_{ij}}
{\overline{\widehat\eta}_{ij}}\quad \mathbf {v_{ij}}\cdot \mathbf {s_{ij}}<0\cr\cr
0\qquad\qquad\qquad\quad \mathbf {v_{ij}}\cdot \mathbf {s_{ij}}>0\cr}
\label{eq:eq24}
\end{equation} 

\noindent 
being $\mu_{ij}$: 

\begin{equation}
\mu_{ij}=\frac{h_{ij}~\mathbf{v}_{ij}\cdot~\mathbf{s}_{ij}}{s^2_{ij}+\nu^2}
\label{eq:eq25}
\end{equation}

Expressions (22),(23) and (24) account for the cartesian part of viscosity. It 
has been shown (i.e. 
Monaghan 2005) that, in the continuous limit, these expressions become 
the Navier-Stokes equations provided shear and bulk viscosity coefficients 
are taken $\eta_{vis}=\rho\alpha_0hc/8$~and $\zeta_{vis}=5\eta_{vis}/3$~respectively. However, in cylindric geometry the 
stress tensor which appear in the Navier-Stokes equations also includes a 
term proportional to velocity divergence 
through the 
so called second viscosity coefficient. Therefore an extra term 
containing the magnitude  $(v_r/r)$~must be added to $\Pi_{ij}^{(1)}$~to 
account 
for the convergence of the flux towards the axis. This new 
term, $\Pi_{ij}^{(2)}$, should fulfill a few basic requirements: 1) Far enough from the axis it 
should be negligible, 2) must vanish for those particles with $v_r\ge 0$, 
3) for homologous contractions the viscous acceleration related to that term 
should be negligible, 4) it should be symmetric with respect particles $i$~and $j$~to 
conserve momentum. An expression satisfying the four items is:

\begin{equation}
\Pi_{ij}^{(2)}=\cases
{\frac{-\alpha_1\bar c_{ij} q_{ij}+
\beta_1~q_{ij}^2}
{\overline{\widehat\eta}_{ij}}\quad  v_{r_i} {\mathrm {and}}~ v_{r_j}<0\cr\cr
0\qquad\qquad\qquad\quad {\mathrm {other~cases}}\cr}
\label{eq:eq24}
\end{equation}

\noindent
where $q_{ij}=0.5~(\frac{h_i  v_{r_i}}{r_i}+\frac{h_j v_{r_j}}{r_j})$. The 
resulting viscous acceleration is: 

\begin{equation}
\mathbf a_i^{vis}=\sum_{j=1}^N m_j   
\Pi_{ij}^{2D}\mathbf {D}_i W_{ij}
\end{equation}

\noindent
where $\Pi_{ij}^{vis}=\Pi_{ij}^{(1)}+\Pi_{ij}^{(2)}$
Note that for a homologous contraction $v_r\propto r$~meaning $q\simeq \mathrm{ 
const.}$, hence the gradient of $\left<\eta \Pi^{(2)}\right>$~vanishes fulfilling the 
third requirement above. In diverging shocks $\Pi_{ij}^{(2)}$~vanishes 
and only 
the cartesian part of viscosity matters. Therefore constants $\alpha_0, \beta_0$~should remain close to their standard values $\alpha_0\simeq 1$~and $\beta_0\simeq 2$. All simulations presented in this paper were carried out 
using  $\alpha_1=\alpha_0=1,~ 
\beta_1=\beta_0=2$. In 
converging shocks, however, the effect of $\Pi_{ij}^{(2)}$~is to increase 
artificial viscosity introducing more damping in the system.

Equation (26), has the advantage that is formally similar to that used in 
three-dimensions. Therefore one can benefit from the well known  
features of the  
artificial 
viscosity in 3D,  
 which can be directly translated to the axisymmetric 
version (several useful variations of equation (19) can be found in \citet{mo05}). In  
particular, it is straightforward to write the corresponding energy equation:

\begin{equation}
\left(\frac{du_i}{dt}\right)_{vis}\ =\frac{1}{2}\sum_{j=1}^N m_j   
\Pi^{2D}_{ij}(\mathbf{v_i-v_j})\cdot \mathbf{D}_i W_{ij} 
\end{equation}

\noindent 
which has to be added to the right hand of equation (3).

\subsection{An approach to the conduction term.}

The differential equation describing the evolution of the specific internal 
energy due to conductive or diffusive heat transfer is: 

\begin{equation}
\left(\frac{du}{dt}\right)_{cond}=\frac{1}{\rho}\mathbf\nabla\cdot(\kappa
\mathbf{\nabla} T)
\end{equation}

\noindent
being $\kappa$~the conductivity coefficient which in turn is 
a function of the local thermodynamic state of the material. The main 
difficulty to write 
equation (28) in a discrete equation suitable to SPH calculations is the 
existence of a second derivative. It is well known that second and higher 
order derivatives often pours a lot of numerical noise in disordered 
systems. A way to avoid that shortcoming is to reduce one degree the order 
of the derivative by taking the integral expression of equation (28), 
\citet{br85} 
. It has been shown (see, for example, \citet{ju04})   
that in 3D-cartesian coordinates the following expression allows to approximate a 
second derivative using 
only the first derivative of the interpolation kernel:

\begin{equation}
\left(\nabla^2 Y\right)_i=2\sum_{j=1}^N\frac{m_j}{\rho_j}
\frac{Y_j-Y_i}{r_{ij}^2}\mathbf r_{ij}\cdot\mathbf{\nabla_i W_{ij}}
\end{equation}

\noindent
where $Y$~represents any scalar magnitude and $\mathbf r_{ij}=\mathbf r_i-\mathbf r_j$. An useful algebraic relationship 
which allows to write   
the heat transfer equation in SPH notation is:  

\begin{equation}
\mathbf\nabla\cdot(\kappa\mathbf{\nabla T})=
\frac{1}{2}\left[\nabla^2(\kappa T)-T\nabla^2\kappa+\kappa\nabla^2 T\right]
\end{equation}

Combining equations (29) and (30) the usual form of the heat transfer equation 
used in 3D-SPH studies is obtained. 
A mathematical expression for the heat transfer 
equation in axisymmetric SPH was given by \citet{br03}.   
However in the derivation 
of equation (29) there was not taken into account the $\simeq 1/r$~ term which 
naturally arises whenever the divergence of a vector is estimated in 
cylindric coordinates:

\begin{equation}
\mathbf\nabla\cdot(\mathbf{\nabla} Y)=
\frac{\left(\frac{\partial Y}{\partial r}\right)}
{r}+\frac{\partial}{\partial r}\left(\frac{\partial Y}{\partial r}\right)+
 \frac{\partial}{\partial z}\left(\frac{\partial Y}{\partial z}\right) 
\end{equation}

It will be shown in Section 4.1
how the inclusion of the first term on the right of equation (31)   
improves the quality of the results in a particular test. 

To obtain the adequate numerical approximation to equation (28) in axisymmetric 
SPH we first make use of equation (30):

\begin{equation}
\left(\frac{du}{dt}\right)_{cond}=
\frac{1}{2\rho}\left\{\nabla^2(\kappa T)-T\nabla^2\kappa+\kappa\nabla^2 T\right
\}
\end{equation}

\noindent
Then, using equation (31) to develop each term inside the brackets  
 and 
using the 2D-operator $\mathbf D\equiv(\frac{\partial}{\partial r},\frac{\partial}{\partial z})$ ~instead of $\mathbf\nabla$~the 
following expression is obtained:

\begin{eqnarray}
\left(\frac{du_i}{dt}\right)_{cond}=
\frac{1}{2\rho}\left\{\frac{1}{r_i}\frac{\partial (\kappa T_i)}{\partial r_i}-
\frac{T_i}{r_i}\frac{\partial \kappa_i}{\partial r_i}+
\frac{\kappa_i}{r_i}\frac{\partial T_i}{\partial r_i}\right\}+\cr\cr
\frac{1}{\rho_i}\sum_{j=1}^N \left(m_j\frac{\kappa_i+\kappa_j}{\widehat\eta_j}
(T_i-T_j) \frac{\mathbf s_{ij}}{s_{ij}^2}\cdot\mathbf{D}_i W_{ij}\right)
\end{eqnarray}
 
The magnitudes inside the brackets are the new terms related to the hoop-stress. In order to calculate the derivatives we make use of one of the golden rules
 of 
SPH (\cite{mo05}):

\begin{equation}
\frac{\partial C_i}{\partial r_i}=\sum_j \frac{m_j}{\widehat\eta_j}(C_j-C_i) \frac{\partial W_{ij}}{\partial r_i}
\end{equation}

\noindent
After a little algebra the following expression is obtained:

\begin{eqnarray}
\left(\frac{du_i}{dt}\right)_{cond}=-\pi 
\sum_{j=1}^N m_j\frac{(\kappa_i+\kappa_j)}{\widehat\eta_i\widehat\eta_j}(T_i-T_j)\frac{\partial W_{ij}}{\partial r_i}+\cr
2\pi r_i\sum_{j=1}^N m_j \frac{(\kappa_i+\kappa_j)}{\widehat\eta_i\widehat\eta_j}(T_i-T_j)
\frac{\mathbf s_{ij}}{s_{ij}^2}\cdot\mathbf {D}_i W_{ij}
\end{eqnarray}

The presence of $(T_i-T_j)$~
in the equation ensures that there is not heat flux between different parts of an isothermal system.
Note that the presence of the $r_i$~multiplier in the second term on the 
right side of 
equation (35) does not ensure complete conservation of the heat flux. However 
the total energy losses in the numerical test below simulating a thermal wave 
evolution were negligible. Of course equation (35) can be symmetrized by 
taking the arithmetical mean $\bar r_{ij}$~instead of $r_i$~but in that case the evolution of the thermal
 wave was not so well reproduced. On another note equation (35) is compatible 
to the second principle of thermodynamics in the sense that heat always 
flows from 
high to low temperature particles. To demonstrate this let us take a 
pair of particles $i$~and $j$~so that $T_i>T_j$. The second term on the right
becomes negative because the scalar product 
${\bf s_{ij}\cdot D_i}W_{ij}$~is always negative. 
The first term on the right is negative for $r_i < r_j$~and positive for 
$r_i > r_j$. As the heat flow from particle $i$~must be negative the only 
trouble could come if $r_i > r_j$. Nevertheless, even in that case heat flux 
is still negative if $r_i/(r_i-r_j)>1/2$~because the sign of the second term 
in equation (35) prevails. Thus $r_i/(r_i-r_j)>1/2$~is a sufficient condition 
tho get the right sign of heat flux between any pair of particles. As 
$r_i-r_j\simeq h$~such 
condition is fullfiled in a large domain of the system. The 
exception could be the axis vicinity where $r_i\rightarrow 0$. Nevertheless 
in that case symmetry enforces the heat flux to be negligible.  
Therefore we expect a good behaviour of the heat flux arrow although there 
are not 
excluded marginal violations if resolution is poor and strong heat fluxes were present close to $r=0$. A way to ensure complete compatibility to the 
second principle of thermodynamics is to make zero the flux between any 
pair of particles violating such principle.

\subsection{Self-Gravity}

Current 2D-hydrocodes often handle gravity by solving the Poisson
equation or, if the system remains nearly spherical, by simply
computing the enclosed lagrangian mass in a sphere below the point and 
using the Gauss law. Methods based on the Poisson solvers have proven
very useful to find gravity in eulerian hydrodynamics where the same grid 
used to compute the motion of the fluid elements can be used in the 
calculation of  
gravity. However they have the difficulty to set suitable outer
boundary conditions owing to the infinite range of the gravitational force. 
In lagrangian gridless methods such as SPH it is better to use the direct 
interaction among mass particles themselves to calculate gravity. The
evaluation of the gravitational force through direct particle-particle
interaction leads to an $N^2$ scheme that makes the computation feasible 
only for a limited number of particles. When $N$ is high, as it is frequent 
in three-dimensional calculations, one has to rely in approximate schemes
such as those based in hierarchical-tree methods, \cite{he89}.
However hierarchical methods do not work
efficiently in the 2D-axisymmetric approach because what we call particles are 
in fact rings of different size. Usually the ratio between the radius of these 
rings and the distance to the point where the force needs to be computed 
is too large to permit the  multipolar approach to evaluate the gravitational 
force. 
Fortunately, the good
resolution usually achieved in 2D using a moderate number of particles 
makes
the direct calculation affordable.

 \begin{figure}
 \center
 \includegraphics[scale=0.7]{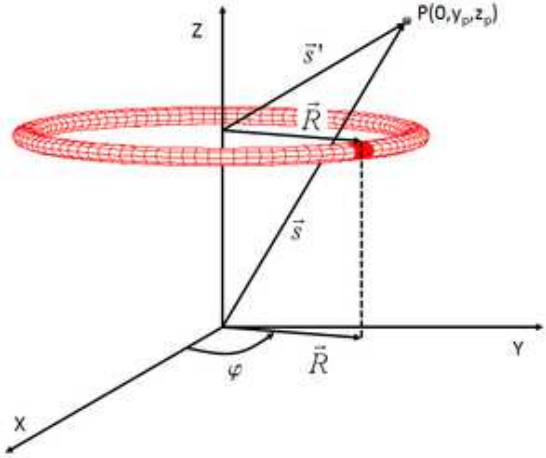}
 \caption{Sketch of the coordinate system and notation used to describe 
gravity.} 
\label{fig2}
\end{figure}

 According to Fig.~2 the gravitational force per unit of mass in a point 
P of coordinates $(0,y_p,z_p)$ (being Z the symmetry axis) due to 
the ring is:

\begin{eqnarray}
\mathbf{g}(0,y_p,z_p)=G\int_0^{2\pi}\frac{\rho R d\varphi}{(R^2+s'^{2}-
2y_p R\sin\varphi)^{3/2}}\times\nonumber \\ 
\left[(R\sin\varphi-y_p)~\mathbf {j}+(z-z_p)\mathbf {k}\right]
\end{eqnarray}

\noindent
where the meaning of the symbols is that shown in Fig.~2. The
gravitational force acting onto the i-particle can be easily written as:

\begin{eqnarray}
\mathbf{g_i}=\sum_{j=1}^N\frac{G}{2\pi}\frac{m_j}{(R_j^2+s_i'^{2})^{3/2}}\times\qquad\nonumber\\
\left[(R_j
I_1-y_i I_2)~\mathbf {j}+(z_j-z_i) I_2)~\mathbf {k}\right]
\end{eqnarray}

\noindent
where $s_i'^{2}=y_i^2+(z_j-z_i)^2$, and $m_j=2\pi R_j\rho_j$ is the mass of
the particle associated to the j-ring. Integrals $I_1$ and $I_2$ are defined
as follows:

\begin{equation}
I_1=\int_0^{2\pi}\frac{\sin\varphi~ d\varphi}{(1-\tau\sin\varphi)^{3/2}}
\end{equation}

\begin{equation}
I_2=\int_0^{2\pi}\frac{d\varphi}{(1-\tau\sin\varphi)^{3/2}}
\end{equation}
where the parameter $\tau$ is:
\begin{equation}
\tau=\frac{2y_i R_j}{R_j^2+s_i'^{2}}
\end{equation}
 Although the elliptical integrals $I_1$ and $I_2$ can not be solved
analytically they can be tabulated as a function of the parameter $\tau$ given
by equation (40). It is straightforward to show that the value of $\tau$~
is always inside the interval $\tau\epsilon [0,1)$, although for $\tau\rightarrow 1$
the integrals $I_1$ and $I_2$ become divergent. Therefore the gravity
force can be calculated in an efficient way
using the following recipe:

          1) Build a table for $I_1$ and $I_2$ as a function of $\tau$. A table
              with $10^4$ rows with values $0\leq \tau\leq 0.9999$ evenly
              spaced is sufficient.

          2) To increase the speed do not interpolate from that table but
              take just the row which is closest to the actual value of $\tau$
              calculated using equation (40).

          3) Note that parameter $\tau$ is symmetric with respect any pair
             of particles, $\tau_i=\tau_j$, thus $I_1(\tau_i)=I_1 (\tau_j)$ and
              $I_2(\tau_i)=I_2 (\tau_j)$. Therefore only half of the 
              interactions
              have to be calculated.

If the algorithm is  optimized the scheme is able to provide the exact
value of the gravity for several dozens of thousand particles. In many
applications using $\simeq 5~10^4$ particles in two-dimensions is enough
to guarantee a good resolution.

Another physical magnitude of interest is the gravitational potential at the
position of the i-particle. It is easy to show that the contribution of the j-
ring to the gravitational potential is:

\begin{equation}
V_i=\frac{G}{2\pi}\frac{m_j}{(R_j^2+s_i'^{2})^{1/2}} I_3
\end{equation}

\noindent
where,
\begin{equation}
I_3=\int_0^{2\pi}\frac{d\varphi}{(1-\tau\sin\varphi)^{1/2}}
\end{equation}

Again the same recipe given above to compute the force can be used
to efficiently calculate $V_i$. On the other hand the exact computation of
the gravitational potential allows to calculate the gravitational force by 
taking the gradient of $V$ at any point.
\begin{equation}
\mathbf {g_i}= -(\mathbf{D} V)_i
\end{equation}

A more suitable form for SPH calculations can be obtained using:

\begin{equation}
\widehat \eta\mathbf{D} V= \mathbf {D}(\widehat{\eta V)}-V\mathbf {D}\widehat\eta
\end{equation}

\noindent
which, according to equations (7) and (9), leads to the following discrete 
equation:

\begin{eqnarray}
\nonumber
\mathbf{g_i}=-(\mathbf{D} V)_i= \frac{f_1^i}{\widehat\eta_i}\sum_{j=1}^N m_j(V_i-V_j)
 \mathbf{D}_i W_{ij}+\\ 
\frac{\left(\frac{df_1^i}{dr}\right)}{\widehat\eta_i}\sum_{j=1}^N m_j(V_i-V_j) W_{ij}~u_r 
\end{eqnarray}

\noindent
where $f_1$~is the corrective term given by equation (5) and $u_r$~the unit 
vector in the r-axis.   

This second route to calculate the gravitational force is computationally
more efficient than evaluating equation (37) because the gradient of 
the potential is
a local quantity which can be calculated in the same part of the algorithm
devised to compute the density or other magnitudes in the hydrocode. It
has the additional advantage that the resulting force is smoothed by the
SPH interpolation procedure avoiding divergences when a pair of particles
become too close.  
In Fig.~5 (bottom-right) there are shown the gravity 
profile  
calculated using equation (45) (filled triangles) and the pressure 
gradient term (continuum line) along a sun-like polytropic structure. As it  
can seen the  
fit is satisfactory except at the surface where the pressure gradient is 
overestimated.   
Although using the potential to calculate gravity is not as exact as the direct  force calculation it is a factor two faster because  
there is a lesser amount of numerical  operations  
in the double loop of 
the gravity routine. 

 Needless to say, the simplicity of the proposed scheme makes the  
parallelization of the gravity computational module straightforward. In this  
case calculations 
with  $10^5-10^6$~particles could become feasible even for  
desktop computers with multiple core processors.

\subsubsection{Free-fall collapse of homogeneous gas structures. Rotation.}

As an initial check of the gravity algorithm resulting from equation (45) we  
have simulated the free-fall collapse of a uniform density sphere of mass 
M$_0$~and radius R$_0$. It is 
a standard test which has the following analytical solution:

\begin{eqnarray}
\frac{t}{t_{ff}}=1-\frac{2}{\pi}\left\{\sin^{-1}\left[\left(\frac{r}{r_0}\right)^{\frac{1}{2}}\right]-\left(\frac{r}{r_0}\right)^{\frac{1}{2}}\left(1-\frac{r}{r_0}\right)^{\frac{1}{2}}\right\}
\end{eqnarray}
  
\noindent 
where $r_0$~is the initial position of the fluid element and 
$0\le t \le t_{ff}$. The free-fall time $t_{ff}$~is:

\begin{eqnarray}
t_{ff}=\frac{\pi}{2}\left(\frac{R_0^3}{2GM_0}\right)^{\frac{1}{2}}
\end{eqnarray}

We built an uniform sphere with  M$_{0}$=1 M$_{\sun}$~and 
R$_0$=R$_{\sun}$~filled with $N=15396$~particles settled in a square lattice. 
Gas pressure and artificial viscosity were set to zero so that the structure 
collapsed under gravity force. Afterwards the implosion  was followed until 
the elapsed  time  was 
close to  
$t_{ff}$. Although restricted to spherical symmetry the free-fall test is 
quite demanding because the evolution is highly non linear, allowing for a 
good check of both the gravity module and the integration scheme (see next section for details). In Fig.~3 there is 
represented the evolution of a particle initially located at 
$r_0=2/3~R_{\sun}$. As we can see its evolution is in good agreement to
 the analytical solution given by equation (46).

 \begin{figure}
 \center
 \includegraphics[scale=0.45]{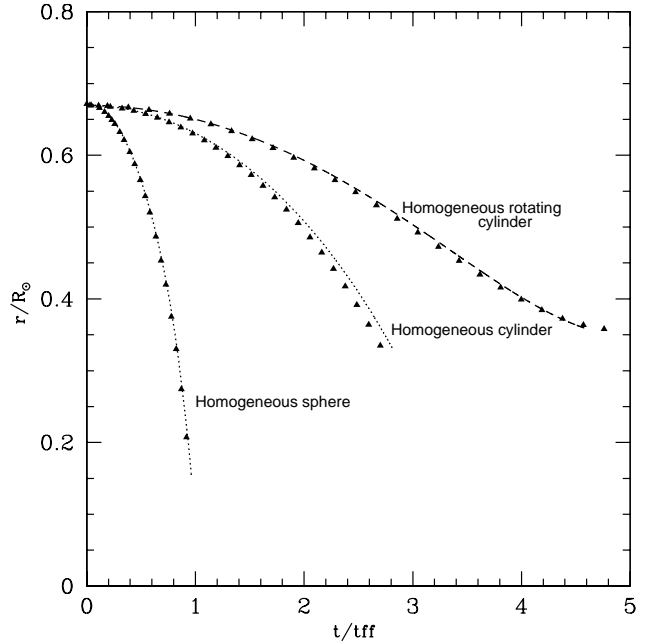}
 \caption{Free-fall test. Triangles represent the analytical solution in each 
case. The initial radius of the particle was $r_0=2/3~R_{\sun}$, $z_0=0$~in 
all cases.  
} 
\label{fig33}
\end{figure}

A question of great interest in astrophysics is the capability of  
axisymmetric SPH codes to handle rotation. The topic is far from trivial
 because in general  
it involves transport of angular momentum via viscous coupling. Even though a 
complete answer to that 
question is beyond the scope of the present work there is a particular case 
that can be handled with the present scheme: the fast implosion (or expansion) 
of a self-gravitating rotating cloud. If the characteristic dynamical time is 
much shorter that the viscous coupling time we can impose angular momentum 
conservation around the symmetry axis to solve this problem. The strategy 
is simply to add the centrifugal 
force which arises from finite angular momentum to the 
r-component of gravity. As an example we have simulated the collapse of a slender rotating cylinder of gas with uniform density centered at the coordinate 
origin. The 
initial conditions are specified by the mass $M_{cyl}$~the radius $R_{cyl}$~and 
the length of the cylinder $Z_{cyl}$. If we suppose {\sl rigid}~rotation 
the specific 
angular momentum of a mass point is given by $L_z(r_0)= w_0~r_0^2$~being 
$w_0$~the initial angular 
velocity of the cylinder and $r_0$~the position of 
the fluid element at the initial time $t_0$. Angular momentum 
conservation demands $L_z(r,t)=L_z(r_0)$~so that the
 centrifugal 
force is $F_c=L_z^2(r_0)/r^3(t)$~which is added to the first 
component of gravity, calculated using equation (45), to obtain
 an effective gravity value. 
As in the case of the spherical collapse
we have taken $M_{cyl}=M_{\sun}$~and $R_{cyl}=R_{\sun}$~while the cylinder
length was $Z_{cyl}=20 R_{cyl}$. The evolution of mass points 
 were followed assuming two values for the angular momentum:
1) zero angular momentum and 2) $L_z (r_0)=\sqrt{(0.5 g_0 r_0^3)}$~where 
$r_0=2/3~R_{cyl}$~for which 
 centrifugal force was a half of gravitational force
at that position. In this case we have used a larger number of particles, 
$N=70000$~  
 to reduce boundary effects. 

On the other hand, assuming $Z_{cyl}>>R_{cyl}$~and using the Gauss-law it is 
possible to work out an analytical approach to the above scenario. The 
acceleration equation becomes:

\begin{eqnarray}
\ddot r=-g_0\frac{r_0}{r}+\frac{L_0^2}{r^3}
\end{eqnarray}

\noindent
where $L_0\equiv L_z (r_0)$~and $g_0$~is the value of gravity at $r=r_0$. According to the Gauss-law 
$g_0=2 G M_{cyl}~r_0/(R_{cyl}^2~Z_{cyl})$,  
but it is better to take $g_0$~directly from the SPH simulation to 
ensure identical initial conditions in both calculations. The solution of equation (48) is:

\begin{eqnarray}
t(r)=-\int_{r_0}^r \frac{dr'}{\left[-2 g_0 r_0 \ln\frac{r'}{r_0}+
 L_0^2\left(\frac{1}{r_0^2}-\frac{1}{r'^2}\right)\right]^{\frac{1}{2}}}
\end{eqnarray}

\noindent
which has to be solved numerically once $r_0$~and $L_0$~are specified.  

Fig.~3  depicts the evolution of a particle initially located at 
$r_0=2/3~R_{cyl}, Z_0\simeq 0$~without and with 
initial angular momentum as well as that obtained using the the analytical 
approach, 
equation (49). The case with zero angular momentum led to the free fall 
of the mass element which was less violent than the spherical case  
owing to the lower initial density in the cylinder. As it can be 
seen in  Fig.~3 the   
agreement between analytical and SPH results is not as good as in the spherical case for $t>t_{ff}$. This is not surprising because boundary effects at
 cylinder edges progressively affects gravity at current particle test 
position and its evolution is very sensible to small variations of gravity 
force. 

When angular momentum was added to the cylinder at 
$t_0=0$~s the implosion of the structure slowed down. As commented above 
the amount of angular momentum  
was chosen to get a centrifugal force contribution at $r_0=2/3~R_{cyl}$~ 
equal to a half of gravity force at that position. As we can see in Fig.~3 
the 
result of 
the simulation is in better agreement to the analytical approach than the 
pure free-fall case. This is because the evolution is driven not only by 
gravity and errors due to the finite size of the rotating cylinder 
are not affecting so much the outcome as in the non rotating 
case. At $t\simeq 5$~t$_{ff}$~the fluid element begins to be centrifugally sustained, in good match to the 
analytical estimation.

\section{Test cases}

In this section we describe and discuss in detail five test cases addressed to validate the 
computational scheme developed in the precedent sections. The first test 
involves the propagation of a thermal discontinuity born at the symmetry axis. 
It is a well known calculation which has an analytical solution to compare 
with. 
Three calculations: the gravitational collapse of a polytrope,  
 the implosion of a homogeneous capsule,  and the wall shock problem deal 
with implosions 
of spherically symmetric systems. Although, at first glance, such constraint looks too restrictive 
in fact it is not, because the spherical symmetry is not a natural 
geometry for  
axisymmetrical systems described with cylindrical  coordinates. Any deviation from the pure 
spherical symmetry during the implosion will trigger the growth of hydrodynamical instabilities. 
Therefore the preservation of the symmetry during the calculation is an important feature 
to be added to  mass, momentum and energy conservation. Finally the last test 
was devoted to simulate the collision of two bubbles of fluid along the Z-axis.

The initial models were calculated by mapping the spherically symmetric 
profiles into a 2D distribution of particles settled in a square lattice. 
The mass of the particles was conveniently adjusted to reproduce the 
density profile of the one-dimensional model. For example, masses
 proportional 
to the $r-$coordinate of the particle were taken to obtain models with 
constant density. Reflective boundary conditions using specular ghosts 
particles 
were imposed across the Z-axis. 
The EOS was that of an ideal gas 
with $\gamma=5/3$~and radiation in the test devoted to the collapse of a 
polytrope, and only gas ideal for the other tests. We have used and 
adaptive smoothing 
length parameter $h(\mathbf s, t)$~which was updated at each time step to 
keep a constant number of neighbours $n_v=36$~within a circle of radius $2h$. 
The interpolating kernel was the cubic polynomial spline. Numerical integration of 
SPH fluid equations was performed using a two-step centered scheme with second 
order accuracy. It is worth to mention that in the free-fall test of a 
homogeneous sphere the velocity was updated using the XSPH variant of \citet{mo89}
to avoid particle penetration through the symmetry axis. In this particular 
test 
mass points at low radius are proner to cross Z-axis because pressure and viscous forces were 
artificially set to zero. Therefore any residual value of gravity  
can impel particles located close to the center towards unphysical 
$r<0$~values. 
We have updated only the $r$-component of velocity using the 
following expression:

\begin{eqnarray}
\tilde{\dot r_{i}}=\dot r_{i}+\epsilon~2\pi~r_i \sum_{j=1}^N\frac{m_j}{\widehat\eta_{ij}}\left(\dot r_{j}-\dot r_{i}\right)~W_{ij}
\end{eqnarray}
  
\noindent
where $\tilde{\dot r_i}$~is the new, smoother, velocity.  
The parameter $\epsilon$~was taken variable:

\begin{equation}
\epsilon=\cases
{\epsilon_0\left(1-\frac{\left(\frac{r}{h}\right)^2}{9}\right) 
 \qquad\qquad r\le 3h \cr
 0 \qquad\qquad\qquad  r > 3h\cr}
\label{eq:eq24}
\end{equation}

\noindent
where $\epsilon_0=0.5$. Using XSPH variant to move particles enforces  
 $\dot r\rightarrow 0$~when $r\rightarrow 0$, making it difficult for 
particles to cross the axis.

\subsection{Evolution of a thermal discontinuity}

Lets consider the problem relative to the propagation of a thermal wave 
moving through an uniform static medium. This is a well known test, addressed 
to check the capability of the numerical scheme to handle thermal 
discontinuities. The initial model was a sample of 57908 particles evenly 
distributed in a square lattice so that the density was $\rho=1$~\den. A 
$\delta$-like jump in internal energy originates a 
a thermal wave front which   
evolves according to:

\begin{eqnarray}
u(r,z,t)=\frac{A}{(4\pi c_v k t)^{\frac{3}{2}}} 
\exp\left(-\frac{r^2+z^2}{4 c_v k t}\right) + u_0 
\end{eqnarray}
  
\noindent
where $c_v$~is the specific heat and $k$~is the thermal conductivity. The 
following set of values were taken:  $A=10^5$~ergs.cm$^3$/g, $u_0=10^3$~erg.g$^{-1}$ and 
$c_v k=1$~cm$^2$.s$^{-1}$. The initial internal energy profile was that given 
by equation (52) for $t=1$~s, which was taken as the initial  
time (t=0 s) for the 
SPH simulation. The evolution of the thermal signal is then controlled by 
the heat conduction equation, equation (35). 

 \begin{figure*}
 \center
 \includegraphics[scale=0.7]{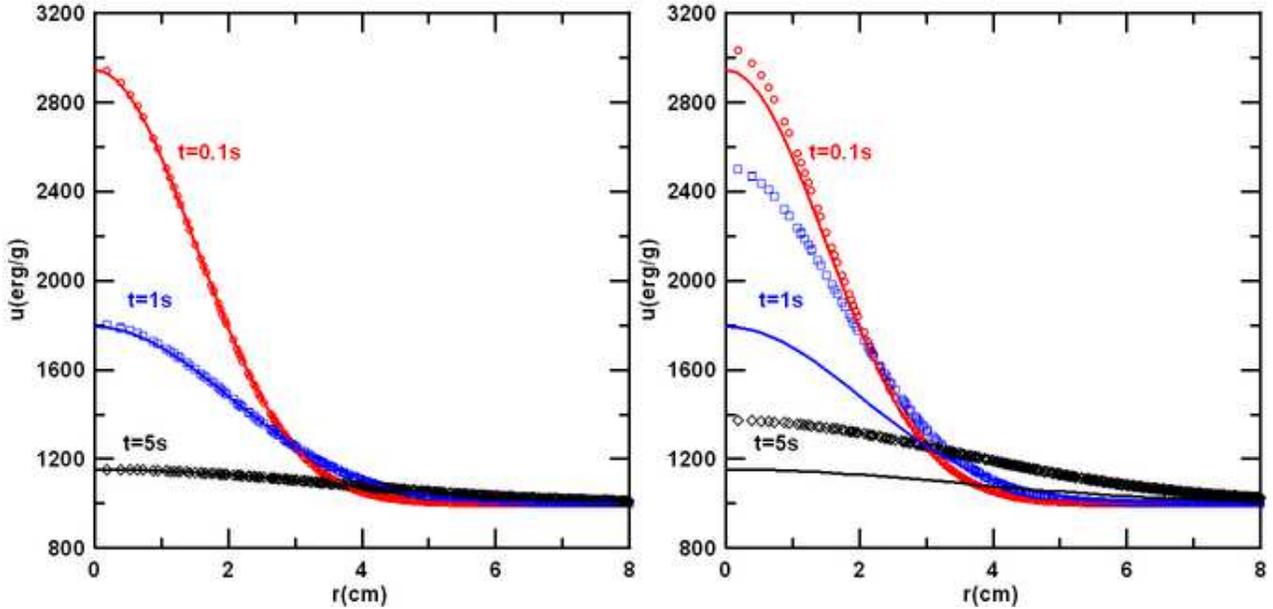}
 \caption{Evolution of a thermal discontinuity initially seeded around the 
symmetry axis. The profile of the internal energy at different times is well 
    reproduced 
   by the SPH when the hoop-stress term is included (left). However when that 
   term 
   is neglected the result does not match the analytical solution represented 
   by the continuum line (right).} 
  \label{fig3}
\end{figure*}

In Fig.~4 (left) there is represented the thermal profile at different 
times. As 
we can see the coincidence with the analytical solution given by equation (52)
 is 
excellent. As time goes on the peak of the signal and its slope decreases  
due to heat diffusion. The initial discontinuity is rapidly 
smeared out by thermal diffusion and soon a thermal wave is born which 
travels to the right, equalizing the internal energy of the system. At t=5 s 
the profile of the internal energy of the gas 
is already quite flat and the system is not far from thermal equilibrium. At  
that moment total energy was conserved up to $5~10^{-5}$.  
In Fig.~4 (right) it is shown the evolution of the thermal profile 
which results when the first term on the right side in equation (35) is removed. As we 
can see the evolution is no longer reproduced by the SPH calculation. 
Therefore it is of utmost importance to include that term, especially in 
those calculations dealing with strong thermal gradients close to the 
symmetry axis. 

\subsection{Gravitational collapse of a polytrope}

The second test involves a catastrophic, albeit highly unprobable, 
astrophysical 
scenario. A spherically symmetric sun-like polytrope  was 
suddenly unstabilized by removing the $20\%$~of its internal energy so that 
the structure collapsed under the gravity force. At some point the collapse is 
halted and an accretion shock forms which manages to eject part of the mass of 
the polytrope. Several episodes of recontraction followed by mass loss ensued 
until the star sets in a new equilibrium state. Even though that particular 
scenario is not realistic it contains several pieces of physics 
of great interest because accretion shocks and pulsational 
instabilities are ubiquitous in astrophysics. As the initial model has 
spherical 
symmetry we expect it to be preserved during the implosion and 
further bounce. The conservation of the symmetry is a demanding test for 
multidimensional hydrocodes in those cases where there are episodes of 
strong decelerations. In the particular case of axisymmetric hydrodynamics 
the higher numerical noise close to the symmetry axis may trigger 
the  growth of convective instabilities. An additional advantage of considering an spherically symmetric initial model is that the evolution calculated with 
the SPH code can be checked using  
standard lagrangian hydrodynamics in one dimension. 

 \begin{figure*}
 \centering
 \begin{minipage}{140mm}
 \includegraphics[scale=0.70]{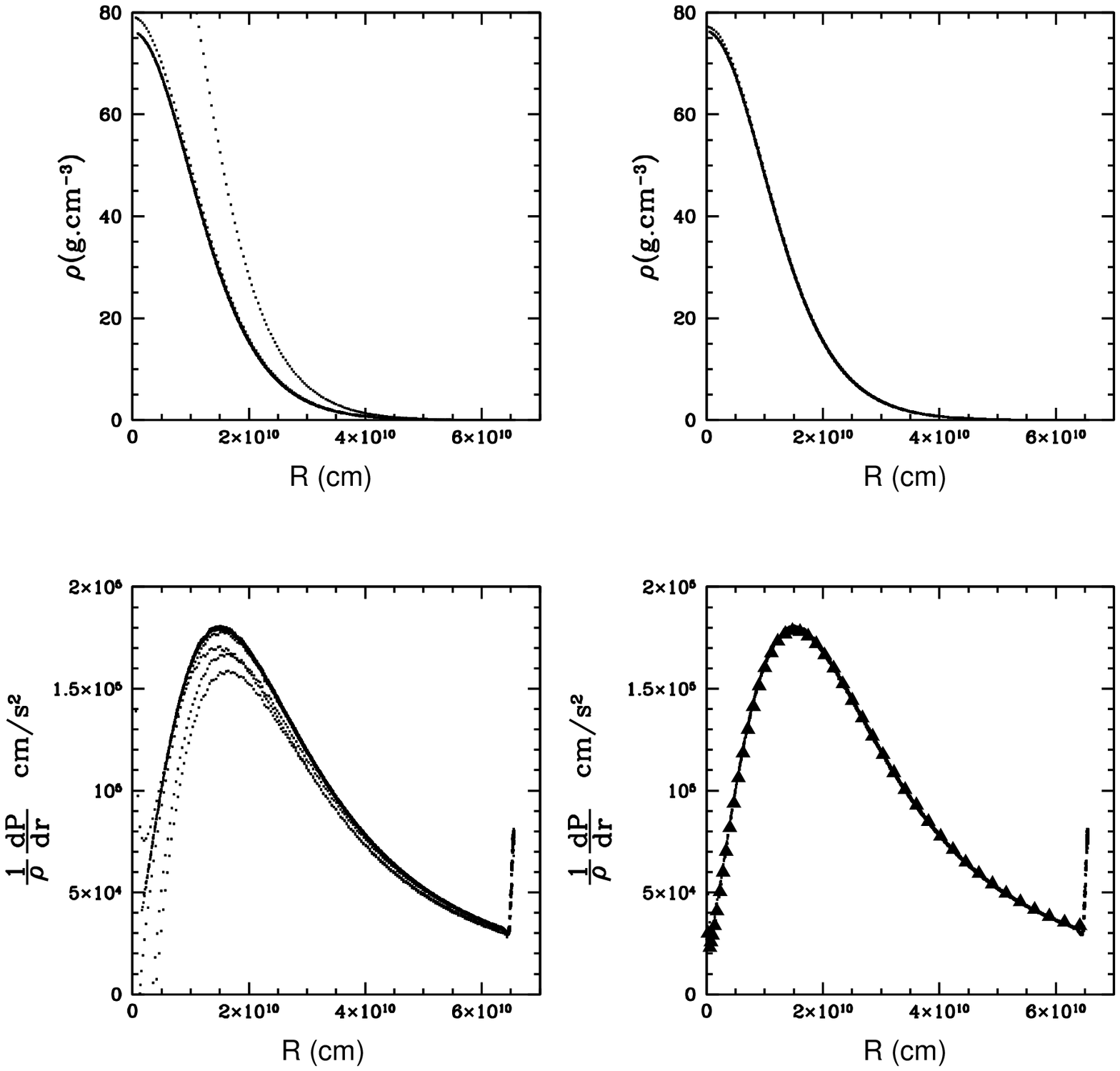}
 \caption{Density, gravity and pressure gradient profiles of the polytrope. 
          Upper-left and right: density profile without and with axis 
          corrections respectively.   
          Bottom-left and right: the same but for the pressure gradient. The 
          absolute value of gravity computed using the gradient of the 
          gravitational 
          potential calculated using equation (45) is marked with full 
          triangles (bottom-right). All mass points of the polytrope have 
          been represented. Neglecting axis corrections leads to a 
          much larger dispersion in the profiles.  
  }\label{fig4}
\end{minipage}
\end{figure*}

The initial model was a $1$~M$_{\sun}$~spherically symmetric polytrope of 
index $n=3$. The radius was set equal to $1$~R$_{\sun}$~so that the 
central  density was 
$\rho_c=77$~\den~. Once the 1D equilibrium model was built it was mapped 
to a 2D distribution of 51408 particles located in a rectangular 
lattice. The mass of the particles were conveniently adjusted to reproduce 
the density profile of the polytrope.
In Fig.~5 there are shown the profiles of density and 
gradient of pressure at t=0 s calculated using the 2D-SPH. As we can see 
the inclusion of the corrective term $f_1$~given by equation (5) in the 
momentum 
equation is crucial to get good enough profiles of these quantities to 
guarantee 
the stability of 
the initial model. The polytrope was perturbed by reducing the 
temperature everywhere in a 20\% of its equilibrium value. Afterwards the 
evolution was followed with the 2D-SPH from the implosion until the first 
pulsation and compared to that obtained by a 1D lagragian hydrocode. 
The main features of the model and a summary of the results are shown 
in Table 1. 

\begin{table*}
 \centering
 \begin{minipage}{140mm}
  \caption{Main features of test models described in Sections~4.2 and 4.3. 
    Conservation of momentum is given by the center of mass displacement 
$\Delta\bar r=\left[(\bar r (t)-\bar r_0)-\bar v_r^0 t\right]$~
and $\Delta\bar z=\left[(\bar z (t)-\bar z_0)-\bar v_z^0 t\right]$~ 
    divided by the radius of the configuration at that time (columns 5 and 6). Momentum  and 
    energy 
    conservation correspond to the last calculated model shown 
in Figs.~6, 7, 8 and 9. }

  \begin{tabular}{@{}llrrrrlrlr@{}}
  \hline
   Test & Number of Part. & $\left(\frac{\rho_{max}}{\rho_0}\right)_{1D}$
 \footnote{Analytical value for the Noh test}& $\left(\frac{\rho_{max}}{\rho_0}\right)_{2D}$
     & $\frac{\vert\Delta\bar r\vert}{R}
$ & $\frac{\vert\Delta\bar z\vert}{R}$&$\frac{\vert\Delta E\vert}{E_0}$\\
 \hline
 \hline
Polytrope  & 51408&  2.43   &2.49   & $5~10^{-15}$ & $10^{-11}$ & 
$5~10^{-3}$ \\
 Capsule implosion & 30448& 29& 32 & $4~10^{-14}$ & $10^{-11}$&$8~10^{-3}$ \\
Noh test & 50334& 64& 58 & $6~10^{-14}$ & $3~10^{-8}$&$2~10^{-3}$ \\
Gas clouds collision & 55814& - & - & $5~10^{-14}$ & $1.3~10^{-3}$&$10^{-2}$ \\
\hline
\end{tabular}
\end{minipage}
\end{table*}

Soon after the model was destabilized the polytrope started to collapse. At 
t$=960$~s a maximum of the central density $\rho_{max}=192.8$~\den was reached. 
A similar maximum of $\rho_{max}=187.1$~\den was obtained using  one-dimensional hydrodynamics. 
The profiles of density and radial velocity at different 
times are depicted in Fig.~6. As we can see the evolution calculated in  
one and two dimensions is very similar and, in general, both profiles are 
in good 
agreement. At our last calculated time, t$=1545$~s, the shock is already 
breaking out the surface of the polytrope. Shortly after that time some 
mass is   
ejected from the surface and, as the 1D calculation shows, the star 
embarks in a long pulsational stage until a new equilibrium state is achieved.

Therefore the numerical scheme was able to handle with this scenario. The 
algorithm devised to calculated the gravity using equation (45), which relies 
in the direct 
interaction between rings (Fig.~2), did a good job. The artificial viscosity 
module was 
also able to keep track with the shocks although, at some stages, the 
post-shock 
region showed a small amount of spurious oscillations. These numerical 
oscillations in the velocity profile are clearly 
visible in Fig.~6 at t$=1545$~s. In the last three columns of  Table 1 
there is information concerning momentum and energy conservation at t$=1545$~s, our last 
calculated model. There was an excellent 
momentum conservation, close to machine precision, whereas the conservation of 
energy was more modest, $\simeq 0.5\%$. On the other hand the spherical 
symmetry was also 
quite well preserved during the calculation. On the negative side we find that 
the two-dimensional calculation was systematically delayed with respect 
its one-dimensional counterpart. The relative shift in time  
remained 
approximately constant, around $1,5$\%, during the evolution. For the sake of 
clarity the elapsed times shown in Fig.~6 were that of the SPH simulation 
and the 
times of the one-dimensional evolution were conveniently chosen to fit the 
2D profiles. 

 \begin{figure*}
 \centering
 \includegraphics[scale=0.7]{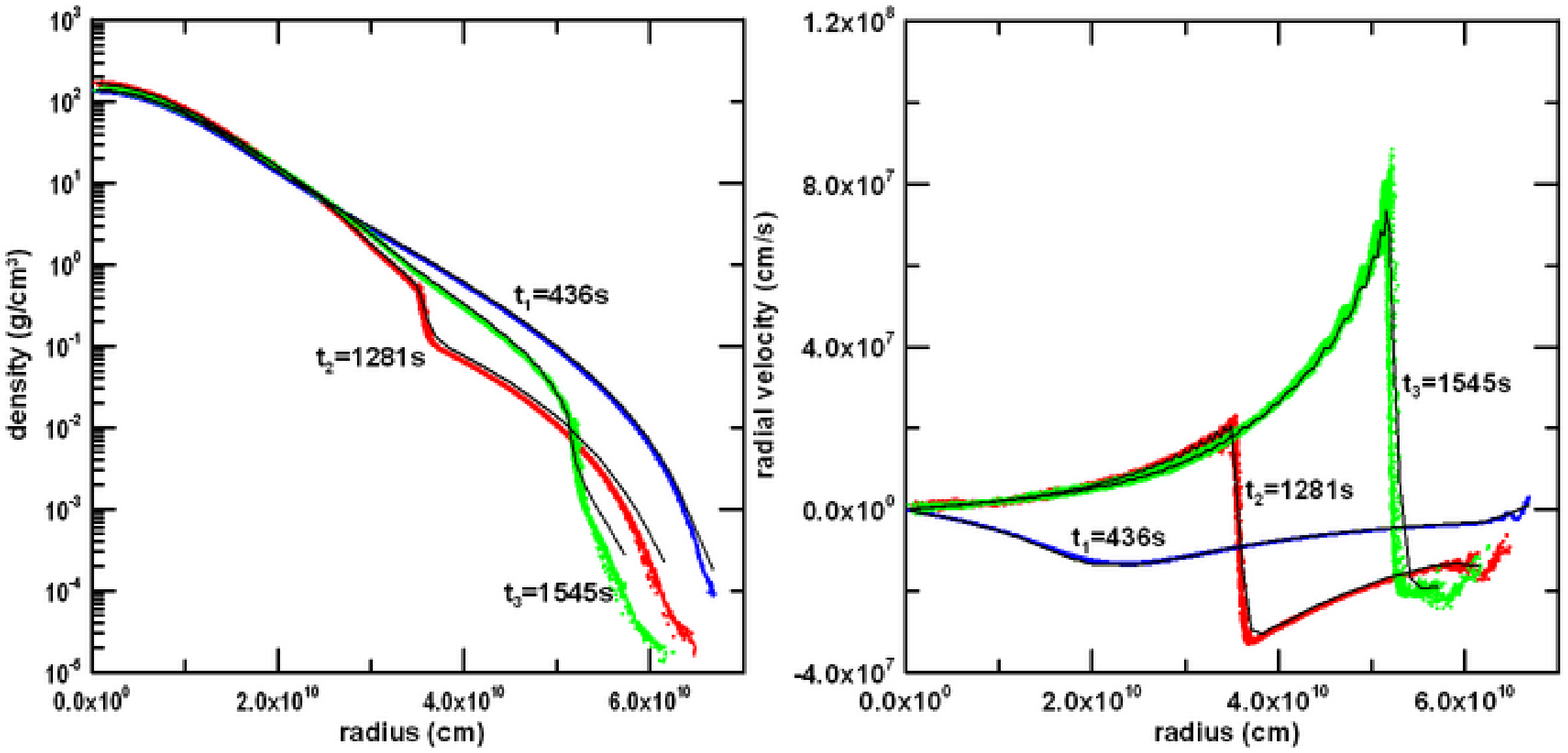}
 \caption{Density profile (left) and radial velocity (right) 
         during the implosion and further rebound of the polytrope at 
         different times. The black continuum line is the profile calculated 
         using a one-dimensional hydrocode of similar resolution. At t=1545 s 
         the shock wave is breaking the surface of the star. In the 
         figure all particles 
         used during the calculation are shown.
  }\label{fig6}
\end{figure*}

\subsection{Implosion of a homogeneous capsule}

Probably the most unfavourable case to numerical simulation is that of a strong
implosion as it currently appears, for example, in ICF studies. Thus, our third test deals 
with the implosion of a homogeneous spherical capsule of size $R_0=1$~ cm and 
density $\rho_0=1$~\den, induced 
by 
the (artificial) 
ablation of its surface. The ablation of the capsule was triggered by the 
instantaneous deposition of energy in the outermost layers of the capsule, which 
had their internal energy increased in the same proportion. 
The energy deposition profile was taken linear from $s=0.8$~cm to $s=R_0=1$~cm so 
that 
the internal energy at $R_0=1$~cm was a factor $10^4$~higher than that at s=0.8 cm.
Below s=0.8 cm a flat profile of the internal energy was assumed. 
The rocket effect caused by the evaporation of the surface layers forms  
a strong shock wave which compresses the interior of the capsule.
The convergence of the shock at the center of the sphere increases the 
temperature and the density in a large factor, much larger than that obtained  
in the previous section dealing with the collapse of a solar-like polytrope. The main 
features of the model are shown in Table 1 and Fig.~7. The profiles of 
density and radial velocity at several times are depicted in Fig.~7 which 
also shows the profiles obtained using a 1D-lagrangian 
hydrocode (continuum lines) with the same physics and identical initial 
conditions.  
The evolution of the capsule can be summarized as follows. Soon after the 
initial energy deposition a pair of shock waves moving in opposite directions 
appear   
(profiles at t=0.0044 s in Fig.~7). As the reverse shock 
approaches the center it becomes stronger owing to the spherical convergence 
(profiles at t=0.0087 s in Fig.~7). Once the maximum 
compression point is reached, $\rho_{max}=31$~\den at t$=0.0111$~ s, the wave 
reflects. When t=0.0150 s the density peak 
has already dropped to 7 \den and most of the material of the capsule is 
expanding homologously. At t=0.0264 s the reflected wave reaches the initial  
radius of the capsule, $R_0=1$~cm. At that time the material of the capsule is 
rather diluted $\rho(r) << 1$~\den and the radial velocity profile consist of  
two homologously expanding zones separated by a transition region 
at $s\simeq 1$~cm. At the last calculated time, t$=0.0264$~ s, the outermost 
layer 
of the sphere has expanded until $s=10$~cm, ten times the original 
size of the capsule. 
As we can see the 2D-simulation match quite well the 
1D results. According to Table 1 the central density at the moment of maximum 
compression is almost the same in both calculations. The conservation of 
momentum is very good, close to machine precision (columns 5, 6 of Table 1), while numerical energy losses remained lesser than 
1\% (last column in Table 1). 
However, an inspection of Fig.~7 reveals that the spherical symmetry was not so well preserved as in the  
collapsing polytrope. Now the implosion of the capsule was more violent and  
the 
deceleration phase before the bounce was more intense, making easier the 
growth of instabilities. Even though the initial model had good spherical 
symmetry, the initial distribution of the particles in a regular lattice acts 
as a source of the so called hour-glass instability. Such instability is 
related to the existence of preferred directions along the grid through which 
the stress propagates. As in the previous test another point of conflict between the 1D and the 2D 
calculations is that the 2D evolution is a bit delayed with respect 
its one-dimensional counterpart. For instance, the times at which the maximum 
central densities are achieved are $t_{1D}=0.01095$~s and $t_{2D}=0.0111$~s, so that the 
relative difference was around $1-2$\%. Such percent level of discrepancy 
remained approximately 
 constant during the calculation.

 \begin{figure*}
 \centering
 \includegraphics[scale=0.7]{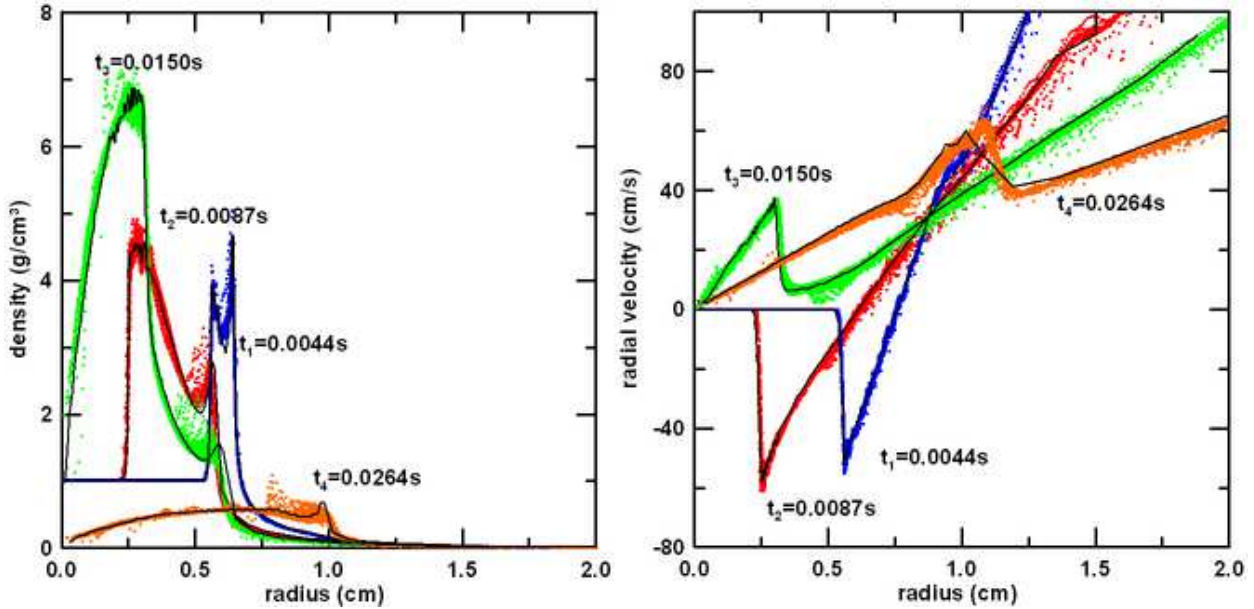}
 \caption{Same as in Fig.~6 but for the small capsule implosion. 
  }\label{fig7}
\end{figure*}

\subsection{Wall heating shock: the Noh test}

The wall heating shock test, \cite{no87}, was especially devised to analyze 
the performace of algorithms addressed to capture strong shocks. Basically 
the wall heating shock test consists in making a sphere or a cylinder implode 
by imposing a converging velocity field. For these geometries there is 
 an analytical approach to the evolution of density and thermodynamical 
variables as a function of the initial conditions. The results of numerical 
codes can be compared with the analytical solution to seek the best method or 
to choose for optimal combination of parameters. It is well 
known that schemes which rely in artificial viscosity have difficulties to 
handle the wall heating shock test. The reason is that artificial 
viscosity 
spreads the shock over a 3-4 computational cells, which induces an  
unphysical rise of 
internal energy ahead the shock. In the case of spherical or cylindrical 
geometry the artificial increase in internal energy is magnified by the  
geometrical convergence of the shock. Therefore the wall shock problem
 represents 
 a strong 
challenge for the axisymmetric SPH. Brookshaw (2003) carried out a similar test 
with the SPH code but far from the symmetry axis. In particular he modelled the 
impact of two supersonic streams of gas, obtaining  good profiles for density 
and internal energy except in a small region around the collision line. The 
density profile showed a dip at that region while a large spike was seen in 
the internal energy. Both, the dip and the spike are numerical artifacts which can be smoothed out by using an artificial heat conduction term to spread 
the excess of internal energy and reduce the 
error around the contact discontinuity. 

To check the performance of our code we have settled N=50334 particles in 
a square lattice. As in the previous tests the mass of the particles was 
conveniently crafted to reproduce an spherical homogeneous system with 
initial radius $R_0=1$~cm.   
The initial conditions were taken as in \citet{no87}: $\rho(s,0)=1$~\den;
 $\dot s(s,0)=-1$~ cm.s$^{-1}$; $u(s,0)=0$~erg.g$^{-1}$. The exact 
solution at time t=0.6 s for $\gamma =5/3$~is shown in Fig.~8 (dashed line). The analytical profile is characterized by a constant post-shock state until 
distance $s=0.2$~cm followed by a rapid decrease in density and internal energy. In the shocked zone density reach a constant value $\rho=64$~\den 
while internal energy was $u=0.5$~erg.g$^{-1}$. The combination of 
such harsh initial conditions 
plus geometrical convergence leads to an strong implosion, even harder 
 than that described in the ablated capsule test. In Fig.~8 (left) there are  
shown the spherically averaged profiles of density and internal energy obtained using SPH at t$=0.6$~s. The error bars in the plot give the 
$1\sigma$~dispersion 
of these variables with respect its mean value in the shell.  
As we 
can see the resulting profiles compare poorly with the analytical results in 
the shocked region. In addition the dispersion is high, especially at low 
radius, a clear signature for the presence of numerical noise. Close to the 
axis there is the typical artificial combination of a dip (in density) and 
spike (in internal energy). The maximum density value was $\rho\simeq 58$~which is around 10\% lower than the exact value. Such bad quantitative 
agreement was not 
unexpected because it is common to all hydrodynamic codes which use the 
artificial viscosity scheme. A way to improve the quality of the simulation 
is to allow for heat conduction in the hydrocode to remove the thermal energy 
spike and to sharpen the shock. Recipes to obtain an artificial conductivity 
coefficient in SPH to better handle the wall shock problem were given by 
\citet{mo92}~and \citet{br03}. For this calculation  we have 
adapted the recipe of 
Monaghan to the features of the axisymmetric SPH defining an artificial 
conductivity for the i-particle:

\begin{eqnarray}
k_i=\bar \rho_{ij}~\bar c_{v_{ij}} \bar h_{ij}\left(\bar c_{ij}+4\vert \mu_{ij}\vert\right)
\end{eqnarray}
  
\noindent
where $\bar c_{v_{ij}}$~is the symmetrized specific heat and $\vert\mu_{ij}\vert$~
is the artificial viscosity parameter given by equation (24). According to  
equation (53) $k_i=k_j$, which can be used directly in equation (35) to compute 
the artificial heat flux. As shown in Fig.~8 (right) the inclusion of the 
artificial heat conduction term leads to a significant improvement of the results. Not only the profound dip in density in the central region has been removed 
but the simulation shows much lesser dispersion around the averaged values of 
density and internal energy. However the maximum peak in density still remains
 $\simeq 10\%$~below the theoretical value. In this respect the only way to 
improve the results is sharpening the shock either by using adaptive kernels,  
\citet{ow98}, \citet{ca08}, or by increasing the number of 
particles.

 \begin{figure*}
 \centering
 \begin{minipage}{140mm}
 \includegraphics[scale=0.35]{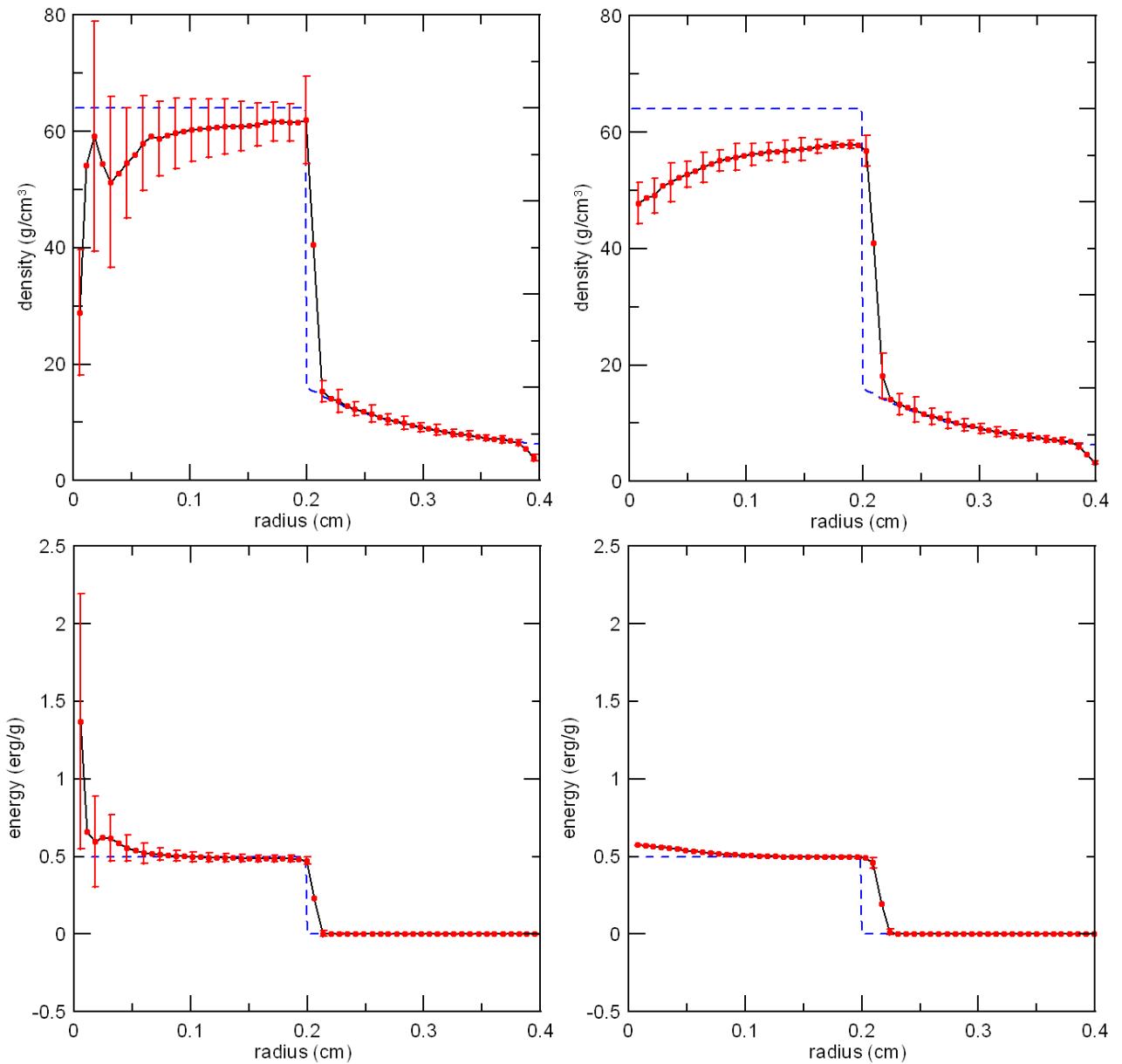}
 \caption{ Results of the Noh test at t=0.6 s. Upper left and right: averaged density 
profiles without and with the artificial heat conduction. Bottom left and right: same but for internal energy. Dispersion around the averaged values are given 
by the error bars. 
  }\label{fig8}
\end{minipage}
\end{figure*}

\subsection{Supersonic collision of two streams of gas}

Our last test is specifically addressed to check momentum conservation in 
a very anisotropic situation: the supersonic collision along the Z-axis of two 
homogeneous spherical 
clouds of gas with different size and mass. The size and mass of the 
spheres are $R_1=1$~cm; $M_1=4\pi/3$~g and $R_2=3$~cm;~$M_2=36\pi$~g 
respectively so 
that their density is $\rho_1=\rho_2=1$~\den. The 
biggest sphere is at rest with its center located at $(0,0)$~cm while the 
smaller one is centered at $(0,-5)$~cm moving upwards with velocity 
$v_z^0=10$~cm.s$^{-1}$. The initial internal energy of both spheres was 
$u_1^0=u_2^0=10^{-3}$~erg.g$^{-1}$~whereas EOS obeys an ideal gas law with 
$\gamma=5/3$. 
The corresponding initial Mach number was  
$M_{\mathrm s}^0\simeq 500$~thus the  
impact is supersonic. The number of particles in each 
sphere was $N_1=5480$~and $N_2=50334$~respectively.  

 \begin{figure*}
 \centering
 \includegraphics[scale=0.5]{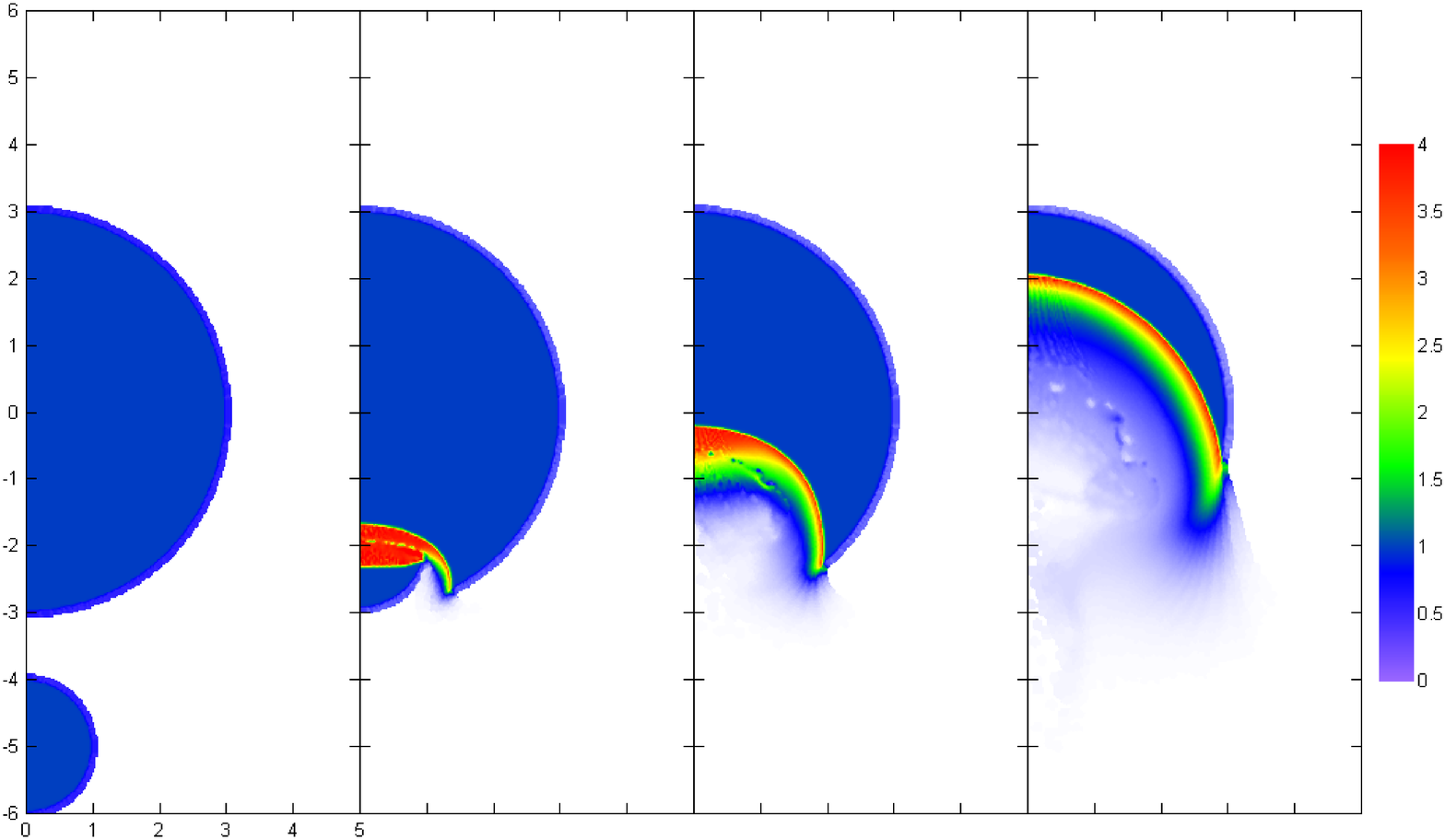}
 \caption{Density colour-map of the collision of two streams of gas at times 
 t$=0$~s, t$=0.31$~s, t$=0.69$~s and t$=1.41$~s. Axis units are in centimeters. 
  }\label{fig9}
\end{figure*}

In Fig.~9 there are represented several snapshots showing the density 
evolution during the collision process. As can be seen the large mass contrast 
leads to the complete deformation of the smaller sphere which, in the end 
transfers most of its initial momentum to the larger bubble. Information  about the evolution of the center of mass of each bubble as well as that 
of the whole system is provided in Fig.~10. According to Figs.~9 and 10 
the collision history can be roughly divided in three stages: 
1) For  $0\le t\le 0.5$~s, 
the incoming smaller cloud deforms while a large fraction of its 
kinetic energy went into internal energy around the collision region. A shock 
wave was launched into the larger bubble, 2) Between 
$0.5\le t\le 1$~s the total internal energy did not change so much. At 
$t=1$~s the velocity of the center of mass of both structures was 
practically the same, 3) For $t>1$~s the energy stored as internal energy is 
again restored to the system. At larger times the velocity of the smaller bubble became  
negative while the bigger cloud acquired a positive velocity 
to preserve total momentum. At our last time $t=1.4$~s the interaction 
between both structures is coming to an end. Fig.~10 also shows that, 
in spite of the large 
changes 
in the velocity of each bubble, the 
velocity of the center of mass of the whole system remained unaltered. 

In the last row of Table 1 there are shown several magnitudes related to momentum and energy conservation. Conservation of linear momentum during the 
interaction was 
monitored through the displacement of the center of mass of the system. At the 
final time, $t=1.4$~s, the deviation of the center of mass position with 
respect the value  
 expected from $\bar v_z^0 t$~was much larger than that shown in the 
other tests. Still the relative error in the position of the center of 
mass once the 
collision has practically ceased was  
 $\simeq 10^{-3}$.  
Angular momentum was very well preserved, 
almost to machine precision. Total energy was conserved to 
$10^{-2}$~relative to the initial kinetic energy of the system. 

In an attempt to understand the origin of the discrepancy relative to momentum 
conservation in the Z-direction we ran exactly the same model but this 
time symmetrizing 
equation (13) (multiplying the term $P_j r_j/\hat\eta_j^2$~ by $f_1^j$). 
There were no significant changes. 
 We conclude that strict total momentum conservation in the Z-direction was
 not possible because of the interaction between  
real and reflected particles. Such interaction took place in a small band 
around the symmetry axis, acting as an external force which modified   
momentum of real particles. However that force can not be balanced by an 
opposite force acting in the left semiplane because, in the Z-direction,
 reflected particles were
 obliged to move exactly 
in the 
same way real particles did. Therefore if strong 
directional anisotropies appear in the vertical displacements of the mass 
points  
some degree of violation in momentum conservation is unavoidable. 
Eventually momentum conservation should improve as the number of 
particles increase because the amount of mass settled in the axis vicinity is 
lower. 

 \begin{figure}
 \center
 \includegraphics[scale=0.4]{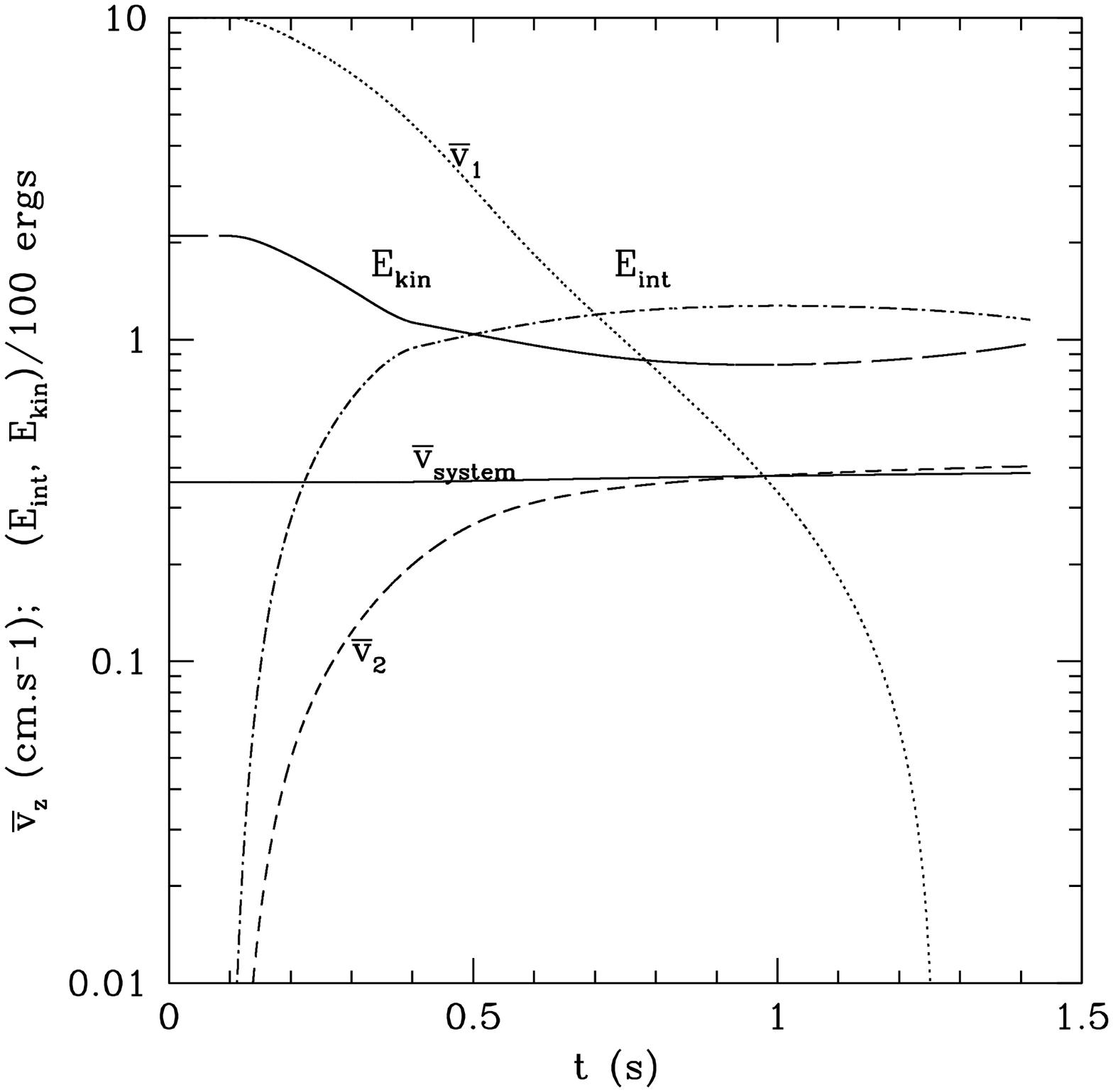}
 \caption{Evolution of $\bar v_z$~of the center of mass of each bubble and  
 of the system. Subscripts 1 and 2 refers to the smaller and larger bubble 
respectively. Evolution of total kinetic and internal energy is also given.} 
\label{fig10}
\end{figure}
  
\section{Conclusions}

Despite the success of the smoothed particle hydrodynamics technique to 
handle gas-dynamics problems in three dimensions little effort has been 
invested to develop two-dimensional (axisymmetric) applications. This work 
intends to fill that gap by solving many of the problems often invoked in 
connection with SPH in cylindric coordinates. These are the 
treatment of the 
singularity axis, the handling of shock and thermal waves and, in many  
astrophysical scenarios, an accurate procedure to calculate the 
gravitational force. 
Our general philosophy in developing the mathematical formalism  was to remain 
as close as possible to the cartesian scheme so that many of the results of 
the standard SPH in three dimensions can be extrapolated with minimum 
changes to the axisymmetric 
version.  

Starting from the fluid Euler equations given by \cite{br03} we have obtained 
analytical corrections to those particles which move close to the singularity 
axis. These corrections appear as multiplicative factors to the different 
terms of the fluid equations. Such multiplicative factors  
become equal to one when $r>2h$~(Fig.~1) so that there is 
no need to 
calculate them beyond that distance and the method is computationally 
efficient.  
Once the 
basic formalism was built we added several pieces of physics which 
make the scheme well suited to handle a large variety of 
problems. First of all an extension of the 3D standard artificial viscosity to 
the 2D-axisymmetric realm was devised, equations (23), (24) and (25). The 
artificial viscosity  
includes the 
 convergence of the flow towards the symmetry axis via  
 linear and  quadratic terms proportional to $[\frac{v_r~h}{r}]$~
(being r the distance to the 
symmetry axis). These terms arise because the diagonal part of the 
stress tensor in cylindric 
coordinates includes the divergence of the flow velocity. However the cartesian 
part of the divergence was already present in the standard formulation given 
by equations (22) and (23) giving rise to bulk and shear viscosity. Therefore 
only the axis converging part of velocity divergence,  
proportional to $v_r/r$~has been added. The resulting viscous force $\Pi^{2D}$~
has two 
terms: $\Pi^{(1)}$~(cartesian) and $\Pi^{(2)}$~(axis converging part) which 
are calculated using similar expressions, equations (23), (25) involving 
four constants $\alpha_0, \beta_0$~($\Pi^{(1)}$)~and $\alpha_1, \beta_1$~($\Pi^{(2)}$). The axis converging part of viscosity will be of importance only for 
strong implosions. Although in the simulations reported in this paper we 
have taken $\alpha_0=\alpha_1=1$~and $\beta_0=\beta_1=2$~there could be other 
plausible combinations worth to explore.  
A similar approach was used to built the conductive transport equation. Starting from the expression given 
by \cite{br85} we have added a new term that 
accounts for the divergence of the temperature gradient in the axis 
neighborhoods. Even though 
the resulting expression was not totally antisymmetric it led to a 
satisfactory energy conservation in the tests described in sections 4.1 and 
4.3. 
 Finally the set of equations was completed with 
the inclusion of self-gravity. In axisymmetric geometry it is better to  
rely  
in the direct, ring to ring, interaction to compute gravity. Such 
procedure, although in general computationally expensive, has several 
advantages: 1) it does not rely in any sort of grid and 
gives the exact value of the gravitational force, 2) the calculations are 
feasible 
for a moderate number 
of particles (i.e. around $5~10^4$~in serial desktop machines) which often 
suffice in 
many 2D simulations  3) the scheme is clear 
and extremely simple making it 
straightforward to parallelize. The algorithm devised to compute gravity 
was able to keep track the free-fall collapse of an homogeneous sphere as 
well as of an homogeneous rotating cylinder. Given an initial 
amount of rotation the evolution of any point of the cylinder was handled
by imposing angular momentum conservation around the symmetry axis. 

Five test cases aimed at  validating the computational scheme were 
considered. Each 
of them intended to check a particular physics item: heat conduction 
(Sects. 4.1 and 4.4), 
shock waves (Sect. 4.3 and 4.4) and gravity (Sects. 3.3.1 and 4.2). Momentum 
conservation was specially checked in Sect. 4.5.
Because for the most part the initial 
configurations had  
spherical symmetry the preservation of that symmetry during the collapse and 
further expansion of the system was taken as an indicator of the general 
ability of the scheme to handle large changes in the spatial scale. 
It was also possible to make detailed comparisons of the results with those 
obtained either 
from analytical calculations, as in the case of the thermal wave propagation, 
or from one-dimensional simulations carried out using standard lagrangian 
hydrodynamics. The results, obtained using several dozen thousand particles, 
were 
in good agreement with the analytical expectations and to the 1D 
simulations. Additionally there was an almost perfect momentum conservation 
while numerical energy losses always remained lesser than $1\%$. The spherical 
symmetry was well preserved although a slight deviation was observed in the 
capsule implosion test, partially due to the so called hour-glass instability 
(owing to the initial setting of the particles in a lattice) and, despite 
the corrective terms, to the 
 effect introduced by the singularity axis onto the approaching particles. 
 A similar but probably more demanding test was the wall heating shock 
problem, 
dealing to the spherically symmetric implosion of a supersonic stream 
of gas. Although the results are not as good as those obtained with other methods which do not 
use artificial viscosity, they are comparable with 
one-dimensional calculations with similar resolution relying in artificial 
viscosity schemes. However the results improved when artificial heat 
conduction, calculated using equations (35) and (53),   
was switched-on.  

As discussed in Sect. 4.5 a weak point of the formulation is that the inclusion of reflective
 particles in the scheme could degrade total momentum 
conservation 
in the Z-direction. However  good 
momentum conservation, much better than  energy conservation for instance, 
 is expected in those systems in which the velocity field is not too 
anisotropic. If strong momentum exchange along the Z-axis is expected,
as in the test described in Sect. 4.5, then momentum will be preserved to 
a similar level as total energy.        

As a conclusion we can say that the formulation of the axisymmetric SPH 
given 
in this paper is a solid tool to carry out simulations 
using that particle method. However there are still a number of unresolved issues 
which deserve further development. One of them is how to build good 
initial models without settling particles of different mass in regular 
lattices.   
This is important because ordered lattices 
introduce spurious instabilities in the system and the mixing of particles 
with 
very different masses could lead to numerical artifacts. Another point of 
difficulty has to do with artificial viscosity, because it introduces 
too much shear viscosity in the system damping the natural development 
of hydrodynamical instabilities. The new term $\Pi_{ij}^{(2)}$~given 
by equation (25) 
of artificial viscosity 
comes from the diagonal of the stress tensor so its contribution 
to shear viscosity is probably lower than that of $\Pi_{ij}^{(1)}$. In 
any case the artificial viscosity formulation 
given in this work is so close to the standard formulation that it could 
 benefit from future advances in the much better studied 
three-dimensional SPH. Finally, the simple test about the implosion of a 
rotating cylinder indicates that axisymmetric SPH is able to handle rotating 
structures but more work needs to be done to incorporate angular momentum 
transport 
into the numerical scheme.

\section*{Acknowledgments}

The authors want to thank the many corrections and suggestions made by the referee. In particular the referee suggested the inclusion of the Noh test and the 
bubble collision scenario described in Sects. 4.4 and 4.5 and inspired the discussion 
about rotation given in Sect. 3.3.1. This work has been funded by Spanish DGICYT grants AYA2005-08013-C03-01 and was 
also supported by DURSI of the Generalitat de Catalunya.

\subsection{Appendices}
\appendix
\section{Correction factors close to Z-axis}

First, we derive the factor $f_1$~affecting the 2D-density $\eta$. 
Close to the Z-axis 
symmetry enforces the 3D-density $\rho$~to have a maximum or a minimum. Thus 
we can safely assume $\rho=\rho_0$~in the axis vicinity. 
If we make the one-dimensional SPH estimation of the averaged density along  
the r-axis:

\begin{equation}
<\eta (r)>=\int\limits_{-\infty}^{\infty}\eta(r') W^{1D} (\frac{\vert r-r'\vert}{h}) dr'
\end{equation}

\noindent 
where $W^{1D}$~is the cubic-spline kernel in 1D:

\begin{equation}
W^{1D}(u_r) = \frac{2}{3h}
\cases
{1-\frac {3}{2}u_r^2+\frac {3}{4}u_r^3\qquad\qquad 0\le u_r\le 1\cr\cr
\frac{1}{4}(2-u_r)^3\qquad\qquad\qquad 1<u_r\le 2\cr\cr
0\qquad\qquad\qquad\qquad\qquad\quad u_r>2\cr}
\label{eq:eq2}
\end{equation}

\noindent
being $u_r=\vert(r-r')\vert/h$. 
If we take  $\eta(r')=2\pi\vert r'\vert \rho_0$, 
which is only valid for $r'\rightarrow 0$:

\begin{equation}
<\eta (r)>=2\pi\rho_0\int\limits_{-\infty}^{\infty} \vert r'\vert 
W^{1D}(\frac{\vert r-r'\vert}{h}) dr'
\end{equation}

\noindent
 Using the cubic-spline the right side of equation (A3) can be integrated to 
give:

\begin{equation}
 <\eta (r)>=\frac{\widehat\eta}{f_1 (\zeta)}
\end{equation}

\noindent
where $\zeta=r/h$~and $\widehat\eta=2\pi r\rho$~is the corrected 
density (hereafter  we put a hat over any corrected quantity). 
The correction factor $f_1(\zeta)$~is given by equation (5). The density in 
brackets is what SPH 
computes using summatories instead of integrals. Thus, adding the z-coordinate 
and using $W^{2D}$~the corrected density can
be evaluated using:

\begin{equation}
 \widehat\eta (r)=<\eta> f_1 (\zeta)=\left(\sum_{j=1}^N m_j W^{2D}(\vert\mathbf{r_i-r_j\vert},h_i)\right)\times f_1^i
\end{equation}

A similar procedure can be used to get the adequate expressions for $\eta v_r$~
and $\eta v_z$, needed to compute the temporal evolution of density, equation 
(15).
Symmetry enforces the radial velocity to vanish at the symmetry axis, thus close to that
 axis $v_r= C r$~and: 

\begin{equation}
<\eta v_r>= \rho (z) 2\pi K^{1D} C\int\limits_{-\infty}^{\infty}\vert r'\vert
 r' W_r(u_r) dr'
\end{equation}

Again the integral on the right admits an analytical solution for the cubic-spline kernel. After a little algebra the
following expression is obtained:

\begin{equation}
<\eta v_r>=\frac{\widehat{\eta v_r}}{f_2(\zeta)}
\end{equation}

\noindent
where $f_2(\zeta)$~is given by equation (16). The expression inside the brackets is what the numerical code calculates using
summatories instead of integrals. Thus, the corrected $\widehat{\eta v_r}$, value of that magnitude is:

\begin{equation}
 (\widehat{\eta v_r})_i=\left(\sum_{j=1}^N m_j v_{r_j} W(\vert\mathbf{s_i-s_j\vert},h_i)\right)\times f_2^i
\end{equation}

Similarly the correction to be applied to $\eta v_z$~is:

\begin{equation}
<\eta v_z>=\frac{\widehat{\eta v_z}}{f_1(\zeta)}
\end{equation}

\noindent
and its corresponding discrete expression is:

\begin{equation}
 (\widehat{\eta v_z})_i=\left(\sum_{j=1}^N m_j v_{z_j} W(\vert\mathbf{s_i-s_j\vert},h_i)\right) \times f_1^i
\end{equation}

A plot of the factors $f_1(\zeta)$~and $f_2(\zeta)$~and their radial derivatives is given  in Fig.~1. The factor $f_1(\zeta)$~is exactly 
 one for $r=2h$ and goes to zero when $r/h\rightarrow 0$, as expected. However, 
it 
can be seen that 
factor $f_2(\eta)$~is close but not exactly one at $r=2h$~and its slope is not 
totally flat. This is because the integral expression in equation (A6)    
is quadratic in the radial coordinate $r$~thus the interpolation does not give 
the exact value of $<\eta v_r>$.

\end{document}